\documentclass[letterpaper,twocolumn,10pt]{article}
\usepackage{usenix-2020-09}
\usepackage{tikz}
\usepackage{amsmath}
\usepackage{acronym}
\usepackage{booktabs}
\usepackage{xspace}
\usepackage{graphicx}
\usepackage{xcolor}
\usepackage[htt]{hyphenat}
\usepackage{comment}
\usepackage{booktabs}
\usepackage{subcaption}
\usepackage{makecell}
\usepackage{cleveref}
\usepackage{booktabs}
\usepackage{multirow}
\usepackage{threeparttable}
\usepackage{makecell}
\usepackage[ruled,vlined]{algorithm2e}
\usepackage{enumitem,kantlipsum}
\usepackage[normalem]{ulem}

\usepackage{kantlipsum} %
\usepackage{import}
\usepackage{footnote}
\usepackage{acronym}
\usepackage{tabulary}
\usepackage{parcolumns}
\usepackage{float}
\usepackage[multiple]{footmisc}
\usepackage{amsthm}
\usepackage{amssymb}%
\usepackage{pifont}%
\usepackage{url}
\usepackage{hyperref}
\usepackage{breakurl}

\newcommand{\keepvspaces}{0} %
\newcommand{\keepvspacesfigure}{1} %
\newcommand{\fname}{Morpheus\xspace}

\newcommand{\cmark}{\color[HTML]{32CB00} \ding{51}}%
\newcommand{\xmark}{\color[HTML]{FE0000} \ding{55}}%

\newcommand\Mark[1]{\textsuperscript#1}

\ifnum\keepvspaces=1
	\newcommand{\vspaceswrap}[1]{\vspace{#1}}
\else
	\newcommand{\vspaceswrap}[1]{}
\fi

\ifnum\keepvspacesfigure=1
	\newcommand{\vspaceswrapfigure}[1]{\vspace{#1}}
\else
	\newcommand{\vspaceswrapfigure}[1]{}
\fi

\usepackage{listings, lstautogobble}
\usepackage{tcolorbox}
\lstdefinestyle{cstyle} {basicstyle=\scriptsize,numbers=left,xleftmargin=2em,frame=lines,framexleftmargin=2em,breaklines=true,moredelim=**[is][\color{red}]{@}{@},moredelim=**[is][\color{green}]{@*}{*@}}

\definecolor{codegreen}{rgb}{0,0.6,0}
\definecolor{codegray}{rgb}{0.5,0.5,0.5}
\definecolor{codepurple}{rgb}{0.58,0,0.82}
\definecolor{backcolour}{rgb}{0.99,0.99,0.99}
\hypersetup{final=true}

\usepackage{tikz}

\input{llvm-ir-lst-def}
\begin{document}

\date{}

\title{\Large \bf Dynamic Recompilation of Software Network Services with Morpheus}

\author{Sebastiano Miano\Mark{1},
        Alireza Sanaee\Mark{1},
        Fulvio Risso\Mark{2},
        G\'{a}bor R\'{e}tv\'{a}ri\Mark{3},
        Gianni Antichi\Mark{1}, \\
        \small{\Mark{1}Queen Mary University of London, UK} \\
        \small{\Mark{2}Politecnico di Torino, IT} \\
        \small{\Mark{3}MTA-BME Information Systems Research Group \& Ericsson Research, HU}
        }

\maketitle

\begin{abstract}
State-of-the-art approaches to design, develop and optimize software packet-processing programs are based on static compilation: the compiler's input is a description of the forwarding plane semantics and the output is a binary that can accommodate any control plane configuration or input traffic.

In this paper, we demonstrate that tracking control plane actions and packet-level traffic dynamics at run time opens up new opportunities for code specialization. We present
\fname, a system working alongside static compilers that continuously optimizes the targeted networking code. We introduce a number of new techniques, from static code analysis to adaptive code instrumentation, and we implement a toolbox of domain specific optimizations that are not restricted to a specific data plane framework or programming language. %
We apply \fname to several eBPF and DPDK programs including Katran, Facebook's production-grade load balancer. We compare \fname against state-of-the-art optimization frameworks and show that it can bring up to 2x throughput improvement, while halving the 99th percentile latency.

\end{abstract}

\acrodef{lpm}[LPM]{Longest Prefix Match}
\acrodef{dwarf}[DWARF]{Debugging With Attributed Record Formats}
\acrodef{jit}[JIT]{Just-In-Time}
\acrodef{ast}[AST]{Abstract Syntax Tree}
\acrodef{ebpf}[eBPF]{extended Berkeley Packet Filter}
\acrodef{xdp}[XDP]{eXtended Data Path}
\acrodef{fdo}[FDO]{Feedback-oriented Optimization}
\acrodef{pgo}[PGO]{Profile-guided Optimization}
\acrodef{acl}[ACL]{Access Control List}
\acrodef{api}[API]{Application Programming Interface}
\acrodef{cpe}[CPE]{Customer Premise Equipment}
\acrodef{dlp}[DLP]{Data Loss/Leakage Prevention}
\acrodef{dpi}[DPI]{Deep Packet Inspection}
\acrodef{dos}[DoS]{Denial of Service}
\acrodef{ddos}[DDoS]{Distributed Denial of Service}
\acrodef{ebpf}[eBPF]{extended Berkeley Packet Filter}
\acrodef{ewma}[EWMA]{Exponential Weighted Moving Average}
\acrodef{foss}[FOSS]{Free and Open-Source Software}
\acrodef{ha}[HA]{Hardware Appliance}
\acrodef{ids}[IDS]{Intrusion Detection System}
\acrodef{ilp}[ILP]{Integer Linear Programming}
\acrodef{isp}[ISP]{Internet Service Provider}
\acrodef{ips}[IPS]{Intrusion Prevention System}
\acrodef{mano}[NFV MANO]{NFV Management and Orchestration}
\acrodef{mips}[MIPS]{Millions of Instructions Per Second}
\acrodef{ml}[ML]{Machine Learning}
\acrodef{nat}[NAT]{Network Address Translation}
\acrodef{nic}[NIC]{Network Interface Controller}
\acrodef{nf}[NF]{Network Function}
\acrodef{nfv}[NFV]{Network Function Virtualization}
\acrodef{nsc}[NSC]{Network Service Chaining}
\acrodef{of}[OF]{OpenFlow}
\acrodef{os}[OS]{Operating System}
\acrodef{pess}[PESS]{Progressive Embedding of Security Services}
\acrodef{pop}[PoP]{Point of Presence}
\acrodef{ps}[PS]{Port Scanner}
\acrodef{qoe}[QoE]{Quality of Experience}
\acrodef{qos}[QoS]{Quality of Service}
\acrodef{sdn}[SDN]{Software Defined Networking}
\acrodef{sla}[SLA]{Service Level Agreement}
\acrodef{snf}[SNF]{Security Network Function}
\acrodef{tc}[TC]{Traffic Classifier}
\acrodef{tor}[ToR]{Top of Rack}
\acrodef{tsp}[TSP]{Telecommunication Service Provider}
\acrodef{vm}[VM]{Virtual Machine}
\acrodef{vne}[VNE]{Virtual Network Embedding}
\acrodef{vnep}[VNEP]{Virtual Network Embedding Problem}
\acrodef{vnf}[VNF]{Virtual Network Function}
\acrodef{vsnf}[VSNF]{Virtual Security Network Function}
\acrodef{vpn}[VPN]{Virtual Private Network}
\acrodef{xdp}[XDP]{eXpress Data Path}
\acrodef{wan}[WAN]{Wide Area Network}
\acrodef{waf}[WAF]{Web Application Firewall}
\acrodef{smartnics}[SmartNICs]{Smart Network Interface Cards}
\section{Introduction}
\label{sec:introduction}

Software Data Planes, packet processing programs implemented on commodity servers, are widely adopted in real deployments \cite{snort,istio,suricata,6379165,hopps:katran,cloudflare:unimog,xhonneux2018leveraging,panda2016netbricks}. This is because they do not require dedicated hardware, guarantee unlimited scale-out/scale-up, and are easier to debug than closed-source hardware~\cite{hopps:katran}. Software data planes depend on a compiler toolchain (e.g., GCC~\cite{gcc} or LLVM~\cite{lattner2004llvm}) to generate machine code, which can be potentially optimized offline through \emph{static} transformations, e.g., inlining, loop unrolling, branch elimination, or vectorization \cite{procieee_2019,alipourfard2018decoupling}. Static optimizations, however, are independent of the actual input the code will process in operation, as this is unknown until then \cite{10.1145/358438.349303, 591653}. Consequently, the resulting \emph{generic} code might contain logic for protocols and features that will never be triggered in a deployment, might be forced to perform costly memory loads to access values that are only known at run time, and take difficult-to-predict branches conditioned on variable data.

\begin{table*}[t]
\centering\renewcommand\cellalign{lc}
\setcellgapes{1pt}\makegapedcells
\scriptsize
\begin{tabular}{l|c|c|c|c|l} 
    \textbf{Name} & \textbf{\makecell{Domain\\specific}} & \textbf{\makecell{Unsupervised\\adaptation to\\control plane\\ actions}} & 
    \textbf{\makecell{Unsupervised\\adaptation to\\data plane\\traffic}} & 
    \textbf{\makecell{Data plane\\agnostic}} & \textbf{Description}\\
    \hline
    Bolt\cite{panchenko2019bolt}        & \xmark & - & - & \cmark & Offline profile-guided optimizer for generic software code.     \\
    AutoFDO~\cite{autofdo}              & \xmark & - & - & \cmark & Offline profile-guided optimizer for generic software code.     \\
    eSwitch~\cite{molnar2016dataplane}  & \cmark & \cmark & \xmark & \xmark & Policy-driven optimizer for DPDK-based OpenFlow software switches.\\
    P5~\cite{abhashkumar2017p5}         & \cmark & \xmark & \xmark & \xmark & Policy-driven optimizer for P4/RMT packet-processing pipelines.\\
    P2GO~\cite{wintermeyer20}           & \cmark & \xmark & \xmark & \xmark & Offline profile-guided optimizer for P4/RMT packet-processing pipelines.\\
    PacketMill~\cite{packetmill}	  	& \cmark & \xmark & \xmark & \xmark &  Packet metadata management optimizer for DPDK-based software data planes.       \\
    NFReducer~\cite{deng2018redundant}	& \cmark & \xmark & \xmark & \cmark & Policy-driven optimizer for network function virtualization. \\
    Morpheus                            & \cmark & \cmark & \cmark & \cmark & Run-time compiler and optimizer framework for arbitrary networking code.
\end{tabular} 
\vspaceswrapfigure{-0.1in}
\caption{A comparison of some popular dynamic optimization frameworks and \fname.} %
\label{tab:optimizations-list-related}
\vspaceswrapfigure{-0.2in}
\end{table*}

\emph{Dynamic} compilation, in contrast, enables program optimization based on invariant data computed at run time and produces code that is \emph{specialized} to the input the program is processing \cite{10.1145/231379.231409, 591653, 10.1145/1542476.1542528}. The idea is to  continuously collect run-time data about program execution %
and then re-compile it to improve performance. This is a well-known practice adopted by generic programming languages (e.g., Java \cite{591653}, JavaScript \cite{10.1145/1542476.1542528}, and C/C++ \cite{10.1145/231379.231409}) and often produces orders of magnitude more efficient code in the context of, e.g., data-caching services~\cite{panchenko2019bolt}, data mining~\cite{autofdo} and databases~\cite{zhang2012micro,kohn2018adaptive}.

To our surprise, we found that state-of-the-art dynamic optimization tools for generic software, including Google's AutoFDO \cite{autofdo} and Facebook's Bolt~\cite{panchenko2019bolt},  
are largely ineffective for network code (\S\ref{sec:motivations}). We demonstrate that the performance of data plane programs critically depends on (i) network configuration, (ii) match-action table content and (iii) traffic patterns, and we argue that standard optimization tools \cite{10.1145/231379.231409, 591653, 10.1145/1542476.1542528} are ill-suited to exploit these \emph{domain-specific} attributes (\S\ref{sec:motivations}). %
Although several tools are available specifically for the networking domain (Table~\ref{tab:optimizations-list}), most perform offline optimizations using recorded execution traces, requiring operators to tediously collect representative samples of match-action tables and predict traffic patterns from production deployments. %
To be practical, instead, a dynamic compiler for networking code should not depend on any \emph{offline} profile, but rather work in a fully \emph{unsupervised} mode where all tracing data needed for code specialization is collected \emph{online}. %
In addition, existing tools are commonly tied to specific hardware, data plane framework, or programming language, limiting their applicability in specific scenarios.

We present \fname, a system to optimize network code at run time using \emph{domain-specific dynamic optimization techniques}. %
\fname operates in a fully unsupervised mode, and it does not require any \emph{a priori} knowledge about control plane configuration or data plane traffic patterns. %
We discuss the main design challenges (\S\ref{sec:architecture}), such as automatically tracking highly variable input (e.g., inbound traffic) that may change tens, or even hundreds of millions of times \emph{per second}. We show that the required profiling and tracing facilities, if implemented carelessly, can easily nullify the performance benefit of code specialization.%

We introduce several novel techniques; we leverage \emph{static code analysis} to build an understanding of the program \emph{offline} %
and introduce a low-overhead \emph{adaptive instrumentation} mechanism to minimize the amount of data collected \emph{online}. Then, 
we invoke several dynamic \emph{optimization passes} (e.g., dead code elimination, data-structure specialization, just-in-time compilation, and branch injection) to specialize the code against control plane actions and data plane traffic patterns. 
Finally, by injecting \emph{guards} at specific points in the pipeline, we protect the consistency of the specialized code against changes to input that is considered invariant (\S\ref{sec:compiler-pipeline}). %

Our implementation, \fname, exploits the LLVM JIT compiler toolchain to apply the above ideas at the LLVM Intermediate Representation (IR) level in a generic fashion and separates data plane specific code to several backend plugins to minimize the effort in porting \fname to a new architecture (\S\ref{sec:implementation}). The code currently contains an eBPF and a DPDK\slash C plugin. We apply \fname to a number of packet processing programs, including the production-grade L4 load balancer Katran from Facebook (\S\ref{sec:evaluation}). Our results show that \fname can improve the performance of the unoptimized (statically compiled) eBPF application up to 94\%, while reducing packet-processing latency by up to 123\% at the 99th percentile. Applying \fname to a DPDK program, we increase performance by up to 469\%.
Finally, we measured \fname against state-of-the-art network code optimization frameworks such as ESwitch~\cite{molnar2016dataplane} and PacketMill~\cite{packetmill}: we show that \fname boosts the throughput by up to 80\% and 294\%, respectively, compared to existing work.

\noindent\textbf{Contributions.} In this paper, we:
\begin{itemize}[leftmargin=0.2in,nosep]
    \item demonstrate that tracking packet-level dynamics opens up new opportunities for network code specialization;
    \item design and implement \fname, a system working with standard compilers to optimize network code at run time;
    \item extensively evaluated Morpheus by applying it to two different I/O technologies (i.e.,DPDK and eBPF), and a number of programs including production-grade software;
    \item share the code in open source to foster reproducibility (\url{link-anonymized}).
\end{itemize}

\section{The Need for Dynamic Optimization}
\label{sec:motivations}
To understand the performance implications of dynamic optimization on software data planes, we present a series of preliminary benchmarks using real network code. We consider two applications: the DPDK sample \emph{firewall} \texttt{l3fwd-acl} \cite{dpdk:l3fwd-acl}, which performs basic L2\slash L3\slash L4 packet processing followed by a lookup into an access control list (ACL) containing a configurable number of wildcard 5-tuple rules, and Katran~\cite{hopps:katran}, Facebook's open-source L4 eBPF/XDP \emph{load balancer}. We connected two servers back-to-back by a 40GbE link, one server running the DPDK \texttt{Pktgen} traffic generator producing a stream of 64-byte packets at line rate \cite{dpdk:pktgen}, and the other running the application under test pinned to a single CPU core (see \S\ref{sec:evaluation} for the details of the configuration).

\begin{figure}[t]
        \centering
        \includegraphics[width=.75\linewidth]{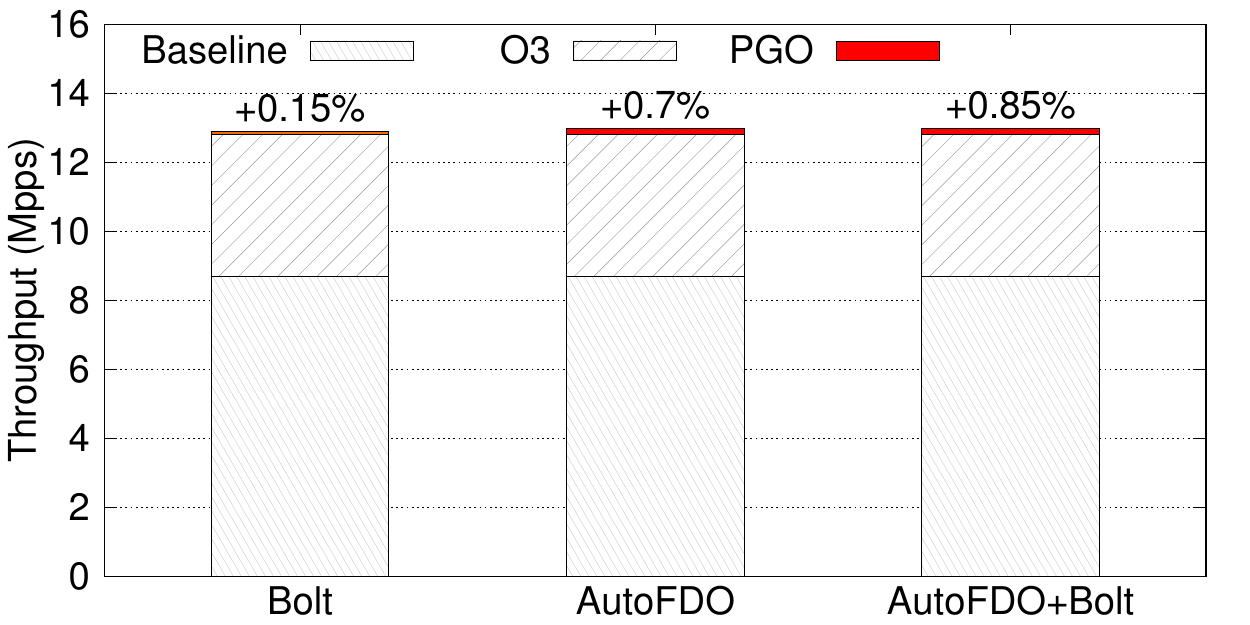}
        \vspaceswrapfigure{-0.1in}
        \caption{Performance breakdown with the DPDK \emph{firewall} application: baseline, enabling level \texttt{-O3} static compiler optimizations, and using various profile-guided optimizations.}
        \label{fig:dyn-opt-generic-pgo}
        \vspaceswrapfigure{-0.2in}
\end{figure}

\vspaceswrap{0.05in}
\noindent\textbf{Generic tools fail to optimize network code.} %
In general, dynamic compilation opens up vast opportunities to improve performance. The question is, to what extent standard dynamic compilers can exploit these, when applied to network software? Fig.~\ref{fig:dyn-opt-generic-pgo} presents the benchmark results for the DPDK \emph{firewall} application at various levels of optimization using standard compiler tools.  In particular, the baseline performance was measured with all optimizations disabled (level \texttt{-O0}), consistently reaching 8.7 Mpps rate in our test. Enabling aggressive GCC \emph{static} optimizations (level \texttt{-O3}, \cite{gcc}) yields $1.5\times$ performance improvement (12.9 Mpps). This is not surprising: it is well-documented that typical C-level static program optimizations greatly benefit networking code \cite{alipourfard2018decoupling, procieee_2019}. %

On top of static optimization, profile-guided optimization tools (PGO), like Google's AutoFDO~\cite{autofdo} or Facebook's Bolt~\cite{panchenko2019bolt}, promise to \emph{dynamically} specialize the code for a specific input by recompiling it based on execution profiles recorded offline.  %
However, our benchmarks indicate that for networking code this promise is not fulfilled; Bolt and AutoFDO could not bring sensible improvements (from $0.15$\% to $0.7$\%), even when the input traffic pattern matches the one used to train the optimization. Using AutoFDO and Bolt \emph{combined} (see~\cite{panchenko2019bolt}), we obtained $1$\% performance increase.

We conclude that generic feedback-based dynamic optimization is mostly ineffective for networking code, as the offline execution profile does not give access to \emph{domain-specific} metrics that are meaningful only in the packet processing context (e.g., match-action table access patterns and table sizes, packet burst size, traffic profiles, or controller configuration).

\noindent%
\noindent\textit{\underline{Takeaway \#1:} Generic dynamic optimization tools fail to optimize typical networking code. This calls for domain-specific dynamic optimizations, which take the specifics of the networking problem space into account.}

\vspaceswrap{0.05in}\noindent\textbf{The promise of domain-specific dynamic optimization.} %
Most data-plane programs are developed as a single monolithic block containing various features that might be activated depending on the specific network configuration in use at any instance of time. For example, many large-scale cloud deployments still run on pure IPv4 and so the hypervisor switches would never have to process IPv6 packets \cite{kubernetes-dualstack} or adopt a single virtualization technology (VLAN\slash VxLAN\slash GRE\slash Geneve\slash GTP) and so switches would never see other encapsulations in operation \cite{ovn-geneve, 6379165}. This implies that, depending on dynamic input that is unknown at compile time, a huge body of unused code gets assembled into the program, boosting code size and causing excess branch prediction misses, negatively impacting the overall performance~\cite{deng2018redundant,liu2018microboxes,miano2018creating, alipourfard2018decoupling}.

\begin{figure}[t]
        \centering
	\includegraphics[width=.75\linewidth]{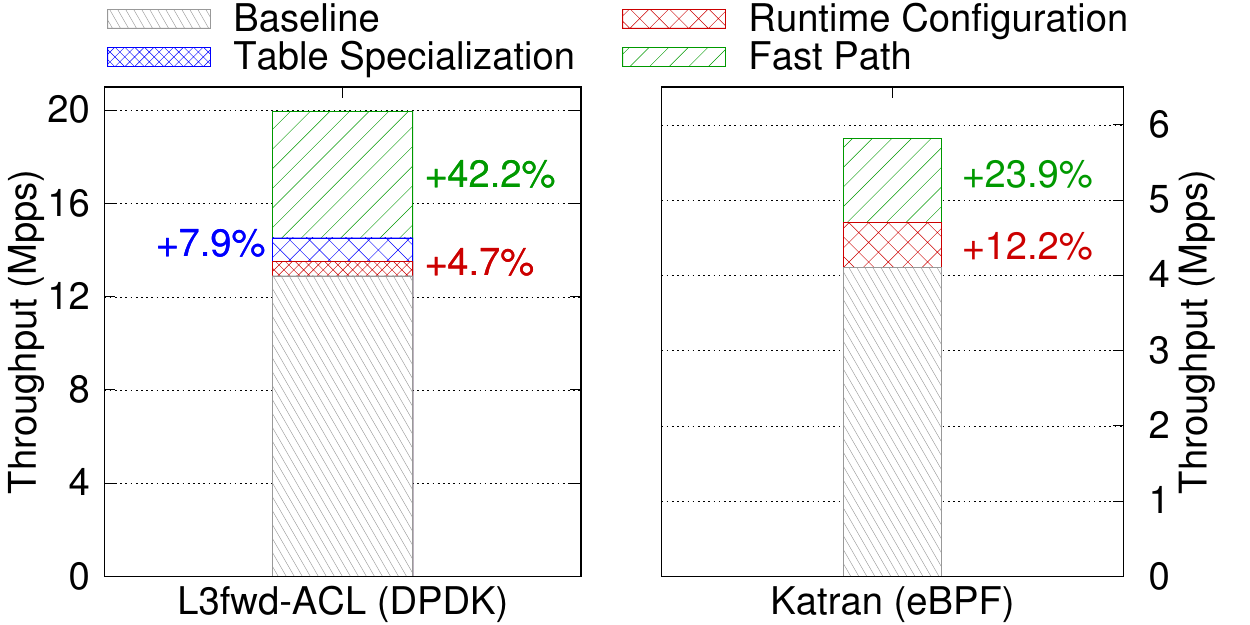}
        \vspaceswrapfigure{-0.1in}
        \caption{Performance breakdown with a set of domain specific optimizations applied to both the DPDK \emph{l3fwd-acl} application and the Facebook's \emph{Katran} eBPF data plane.}
        \label{fig:dyn-opt-generic}
        \vspaceswrapfigure{-0.2in}
\end{figure}

Removing unused code based on \emph{runtime configuration} then can have a profound effect on software performance. To show this, we configured our DPDK \emph{firewall} as a TCP signature-based Intrusion Detection System (IDS), with pure TCP \emph{wildcard} rules generated with ClassBench~\cite{taylor2007classbench}. This opens up a simple opportunity for optimization: all non-TCP packets can now directly bypass the ACL table, avoiding a wasteful lookup. %
Fig.~\ref{fig:dyn-opt-generic} shows the runtime benefit of this optimization (under the \textit{Runtime configuration} bar) for a synthetic input traffic trace where only about 10\% of the input packets are UDP.  Although around 90\% of the traffic still has to undergo an ACL lookup, just avoiding this costly operation for a small percentage of traffic already increases performance with about 4.7\%, without changing the semantic in any way. %

In many practical scenarios, like DDoS blocking, security groups~\cite{openstack,google:k8s} or whitelist-based access control, most firewall rules are fully-specified; for instance, in the official Stanford ruleset~\cite{kazemian2012header} on average \textasciitilde45\% of the rules are purely exact-matching. This opens up another dynamic optimization opportunity: add in front of the ACL an exact-matching lookup table to sidestep the costly wildcard lookup. The result in Fig.~\ref{fig:dyn-opt-generic} (under the \textit{Table specialization} bar) shows a further \textasciitilde8\% performance improvement with this simple modification.

A similar effect is visible with the \emph{load-balancer}: configuring Katran as an HTTP load balancer~\cite{olteanu2018beamer,barbette2020ceetah} allows to dynamically remove all the branches and code unrelated to IPv4\slash TCP processing, which reduces the number of instructions by \textasciitilde58\% (as reported by the Linux \texttt{perf} tool), yielding \textasciitilde17,1\% decrease in the number of L1 instruction cache-load misses. Better cache locality then translates into \textasciitilde12\% performance improvement (from 4.09 Mpps to 4.69 Mpps). %

\noindent\textit{\underline{Takeaway \#2:} Specializing networking code for slowly changing input, like flow-rules, ACLs and control plane policies, substantially improves the performance of software data planes.}

\vspaceswrap{0.05in}\noindent\textbf{The need for tracking packet-level dynamics.} %
The potential to optimize code for specific network configurations has been explored in prior work, for OpenFlow \cite{molnar2016dataplane}, P4 software \cite{retvari2017dynamic,wintermeyer20} and hardware targets \cite{abhashkumar2017p5}, network functions \cite{deng2018redundant}, and programmable switches \cite{packetmill} (see Table~\ref{tab:optimizations-list-related}).  In order to maximize performance, however, we need to go beyond specializing the code for relatively stable runtime configuration and apply optimizations at the packet level.

Consider the DPDK \emph{firewall} application. %
We installed 1000 wildcard rules and generated highly skewed traffic, so that from the thousand active unique 5-tuple flows only 5\% accounts for 95\% of the traffic. This opens up the opportunity to inline the match-action logic for the recurring rules. As the results show (Fig.~\ref{fig:dyn-opt-generic}, under the \textit{Fast Path} bar), we obtain \textasciitilde42\% performance improvement with this simple traffic-dependent optimization. With the eBPF \emph{load balancer} the effect is also visible: configuring 10 Virtual IPs~(VIP) (both TCP and UDP), each with hundred different back-end servers, a similarly skewed input traffic trace presents the same opportunity to inline code, yielding \textasciitilde24\% performance edge. 

\noindent\textit{\underline{Takeaway \#3:} For maximum performance, networking code must be specialized with respect to inbound traffic patterns, despite the potentially daunting packet-level dynamics.}

\section{Challenges}
\label{sec:architecture}
Static compilation performs optimizations that depend only on compile-time constants: it does not optimize variables whose value is invariant during the execution of the program but remain unknown until then. Dynamic compilation, in contrast, enables specializing the code with respect to \emph{invariant} run time data \cite{10.1145/231379.231409}. This opens up a broad toolbox of optimization opportunities, to propagate, fold and inline constants, remove branches and eliminate code never triggered in operation, or even to completely sidestep costly match-action table processing. The \emph{unsupervised optimization of networking code}, however, presents a number of unique challenges:

\vspaceswrap{0.05in}
\noindent\textbf{Challenge~\#1: Low-overhead run time instrumentation.} %
Unsupervised dynamic optimization rests on the assumption that program variables remaining invariant for an extended period of time are promptly detected. The prevalent approach is to collect instruction-level run-time profiles, record the input values and internal variables as well as the associated code execution paths. Then, use this profile to detect hotspots that may be tempting targets for dynamic optimization \cite{10.1145/231379.231409, 591653, 10.1145/1543135.1542528}. However, this is challenging at data-plane time scales: recording at run time instruction-level logs for code that processes potentially tens of millions of packets per second can introduce an overhead that makes the subsequent optimization pointless. We tackle this challenge in \fname by using \emph{static code analysis} to understand the structure of the program offline (\S\ref{sec:ppl-identification}) and leveraging an \emph{adaptive instrumentation} mechanism to minimize the amount of data that is collected online (\S\ref{sec:runtime-stats-collection}).

\vspaceswrap{0.05in}
\noindent\textbf{Challenge~\#2: Dynamic code generation.} %
Once run time profiling information is available, the dynamic compiler applies a set of domain-specific optimization passes to specialize the running code for the profile. Here, \emph{code generation} %
must integrate seamlessly into the compiler toolchain, to avoid interference with the built-in optimization passes. Furthermore, a toolbox of \emph{domain-specific optimization passes} must be identified, which, when applied to networking code, promise significant speedup (\S\ref{sec:dp-optimizations}).

\vspaceswrap{0.05in}
\noindent\textbf{Challenge~\#3: Consistency.} %
The dynamically optimized data plane is contingent on the assumption that the data considered invariant during the compilation indeed remains so: any update to such data would immediately invalidate the specialized code. Here, the challenge is to guarantee data plane consistency under any modification to the invariants on which the specialized code relies. We tackle this challenge by injecting \emph{guards} at critical points in the code that allow the execution to fall back to the generic unoptimized path whenever an invariant changes. %
Since the performance burden on each packet, possibly taking several guards during its journey, can be taxing, we introduce a \emph{guard elision} heuristic to sidestep useless guards (\S\ref{sec:dp-optimizations}). To do so, our static code analysis tool must have enough understanding of the program to separate \emph{stateless from stateful code} (\S\ref{sec:dp-optimizations}). %
Finally, mechanisms are needed to \emph{atomically update} the data plane once the code is re-optimized for the new invariants %
(\S\ref{sec:pipeline-switch}).

\vspaceswrap{0.05in}
\noindent\textbf{Challenge~\#4: Backend independence.} %
Software data planes may run on a diverse collection of backends, including kernel-based virtual machines \cite{hoiland2018express, cilium:ebpf-docs}, kernel bypass frameworks \cite{dpdk:l3fwd-acl}, programmable software switches \cite{openvswitch2015, 8535089} and network function virtualization engines \cite{han2015softnic, barbette:fastclick, panda2016netbricks}.  %
In order to foster portability, the compiler should remain \emph{backend-agnostic} as much as possible.  We tackle this challenge by internally separating out the generic parts of the compiler toolchain into a backend-independent core and hiding backend-specific details behind a versatile backend API.

\section{\fname Compilation Pipeline}
\label{sec:compiler-pipeline}

\begin{figure}[t]
  \centering
  \includegraphics[width=1\linewidth]{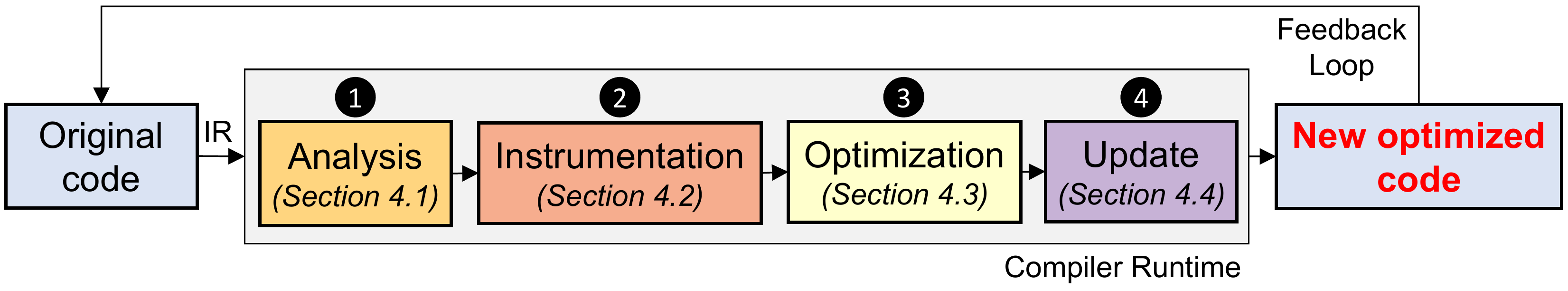}
  \vspaceswrapfigure{-0.3in}
  \caption{The \fname compiler pipeline.}
  \label{fig:overall-architecture}
  \vspaceswrapfigure{-0.2in}
\end{figure}

We designed \fname with an ambitious goal: to build a portable dynamic software data plane compilation and optimization toolbox. The system architecture is shown in Fig.~\ref{fig:overall-architecture}. \fname accepts the input code at the Intermediate Representation (IR) level. The pipeline is triggered periodically at given time slots to readjust the code for possibly changed traffic patterns and control plane updates. At each invocation, the compiler performs an extensive offline code analysis to understand the program control\slash data flow (see \S\ref{sec:ppl-identification}) and then reads a comprehensive set of instrumentation tables to extract run-time match-action table access patterns (see \S\ref{sec:runtime-stats-collection}). Finally, \fname invokes a set of dynamic compilation passes to specialize the code (see \S\ref{sec:dp-optimizations}) and then replaces the running data plane with the new, optimized code on the fly (see \S\ref{sec:pipeline-switch}).

Below, we review the above steps in more detail. We use the simplified main loop of the \textit{Katran} XDP/eBPF load balancer~\cite{hopps:katran} as a running example (see Listing~\ref{lst:katran-ebpf}).
The main loop is invoked by the Linux XDP datapath for each packet. It starts by parsing the L3 (line~\ref{line:katran_l3_headers}) and the L4 (line~\ref{line:katran_l4_headers}) header fields, using a special case for QUIC traffic as this is not trivial to identify \cite{I-D.ietf-quic-manageability}. In particular, QUIC flows are marked by a flag stored in the VIP record (line~\ref{line:katran_quic_vip}); if the flag is set, then a special function is called to deal with the QUIC protocol. Otherwise, a lookup in the connection table (line~\ref{line:katran_conn_lkp}) is done: in case of a match, the ID of the backend assigned to the flow is returned; if no connection tracking information is found, a new backend is allocated and written back to the connection table (line~\ref{line:katran_conn_update}). Finally, the IP address of the backend associated with the packet is read from the backend pool (line~\ref{line:katran_real_lookup}), the packet is encapsulated (line~\ref{line:katran_encapsulate}) and sent out (line~\ref{line:katran_send}).

\subsection{Code Analysis}
\label{sec:ppl-identification}
To be able to specialize code, we need to have a good understanding of the possible inputs it may receive during run time. Networking code tends to be fairly simplistic in this regard: commonly, the input consists of the \emph{context}, which in eBPF/XDP corresponds to the raw packet buffers, and the content of match-action tables named \emph{maps} in the eBPF world (Listing~\ref{lst:katran-ebpf}). %
Since input traffic may be highly variable and provides limited visibility into program operation, 
\fname does not monitor this input \emph{directly} \cite{alipourfard2018decoupling}.  Rather, it relies on tracking the map access patterns and uses this information to \emph{indirectly} reconstruct aggregate traffic dynamics and identify invariants along frequently taken control flow branches. %

\begin{lstlisting}[float,language=c,style=cstyle,escapechar=|,caption={Simplified Katran main loop},captionpos=t,label={lst:katran-ebpf},belowskip=-0.2in]
int process_packet (packet pkt) {
 u32 backend_idx;

 parse_l3_headers(pkt);|\label{line:katran_l3_headers}|
 parse_l4_headers(pkt);|\label{line:katran_l4_headers}|

 vip.vip = pkt.dstIP;|\label{line:katran_vip_dce_start}|
 vip.port = pkt.dstPort;
 vip.proto = pkt.proto;
 vip_info = vip_map.lookup(vip);|\label{line:katran_vip_lookup1}|

 if (vip_info->flags & F_QUIC_VIP){|\label{line:katran_quic_vip}|
    backend_idx = handle_quic();
    goto send;|\label{line:katran_quic_vip_end}|
 }

 backend_idx = conn_table.lookup(pkt);|\label{line:katran_conn_lkp}|
 if(!backend_idx) {
    backend_idx = assign_to_backend(pkt)
    conn_table.update(pkt, backend_idx);|\label{line:katran_conn_update}|
 }
 
send:
   backend = backend_pool.lookup(backend_idx);|\label{line:katran_real_lookup}|
   encapsulate_pkt(backend->ip);|\label{line:katran_encapsulate}|
   return XDP_TX;|\label{line:katran_send}|
}
\end{lstlisting}

In the first pass, \fname uses comprehensive statement-level \emph{static code analysis} to identify all map access sites in the code, understand whether a particular access is a read or a write operation, and reason about the way the result is used later in the code. In particular, \emph{signature-based call site analysis} is used to track map lookup and update calls, and then a combination of \emph{memory dependency analysis} \cite{memoryssa} and \emph{alias analysis} \cite{aliasan} is performed to match map lookups to map updates. Maps that are never modified from within the data plane are marked as read-only (RO) and the rest as read-write (RW). Note that RO maps may still be modified from user space, but such control-plane actions tend to occur at a coarser timescale compared to RW maps, which may be updated with each packet. This observation will then allow to apply more aggressive optimizations to \emph{stateless} code, which interacts only with relatively stable RO maps, and resort to conservative optimization strategies when specializing stateful code, which depend on potentially highly variable RW maps.

\vspaceswrap{0.05in}
\noindent\textbf{Running example.} %
Consider the \textit{Katran} main loop (Listing~\ref{lst:katran-ebpf}). \fname leverages the domain-specific knowledge, provided by the eBPF data-plane plugin (\S\ref{sec:polycube-implementation}), to identify map reads by the \texttt{map.lookup} eBPF helper signature and map writes either via \texttt{map.update} calls or a direct pointer dereference. Thus, map \texttt{backend\_pool} is marked as RO and \texttt{conn\_table} as RW.  For \texttt{vip\_map}, memory dependency analysis finds an access via a pointer (line~\ref{line:katran_quic_vip}), but since this conditional statement does not modify the entry and no other alias is found, \texttt{vip\_map} is marked as RO as well.

\begin{table*}[t]
\footnotesize
\centering
\setlength\tabcolsep{2pt}%
  \begin{tabulary}{1.0\textwidth}{l|l|c|c|c|c}
    \hline
    
    \textbf{Optimization}  & \textbf{Description} & \textbf{Small RO maps} & \textbf{Large RO maps} & \textbf{RW maps} & \textbf{Traffic-dependent}\\
    \hline   
    JIT (\S\ref{sec:jit})			& inline frequently hit table entries into the code & \cmark & \cmark & \cmark & \cmark\\
    Table Elimination (\S\ref{sec:jit})     			& remove empty tables & \cmark & \cmark & \xmark& \xmark\\
    Constant Propagation (\S\ref{sec:constant-prop})		& substitute run-time constants into expressions & \cmark & \cmark & \xmark & \cmark\\
    Dead Code Elimination (\S\ref{sec:dce})			& remove branches that are not being used & \cmark & \cmark & \xmark & \cmark\\
    Data Structure Specialization (\S\ref{sec:ds-spec})	& adapt map implementation to the entries stored & \cmark & \cmark & \cmark & \xmark\\
    Branch Injection (\S\ref{sec:branch-inj}) 		& prevent table lookup for select inputs & \cmark & \cmark & \xmark & \xmark\\
    Guard Elision (\S\ref{sec:guards}) 		& eliminate useless guards & \cmark & \cmark & \xmark & \xmark\\
    \hline	
  \end{tabulary}
  \vspaceswrapfigure{-0.1in}
  \caption{Dynamic optimizations in \fname. Applicability of each optimization depends on the map size, access profile (RO\slash RW), and availability of instrumentation information. Note that optimizations marked as "traffic-dependent" can also be applied, at least partially, without packet-level information (e.g., small RO maps can always just-in-time compiled). For full efficiency, these passes rely on timely instrumentation information (e.g., to JIT heavy hitters from a large map as a fast-path).}
  \label{tab:optimizations-list}
  \vspaceswrapfigure{-0.1in}
\end{table*}

\subsection{Instrumentation}
\label{sec:runtime-stats-collection}
In the second pass, \fname profiles the dynamics of the input traffic by generating \textit{heatmaps} of the maps access patterns, so that the collected statistics can then be used to drive the subsequent optimization passes. %
Specifically, \fname uses a \emph{sketch} to keep track of map accesses, by storing instrumentation data in a LRU (least-recently-used) cache alongside each map and adapting the sampling rate along several dimensions to control the run-time cost of profiling.  The dimensions of adaptation are as follows. %
(1) \emph{Size:} small maps are unconditionally inlined into the code and hence instrumentation is disabled for these maps. %
(2) \emph{Dynamics:} \fname does not record each map access, but rather it samples just enough information to reliably detect heavy hitters \cite{10.1145/510726.510749}. %
(3) \emph{Locality:} instrumentation caches are per-CPU and hence track the local traffic conditions at each execution thread separately, i.e., specific to the RSS context.  This improves per-core heavy hitter detection in presence of highly asymmetric traffic. %
(4) \emph{Scope:} after identifying heavy hitters in the CPU context, local instrumentation caches are run together to identify global heavy hitters. %
(5) \emph{Context:} if a map is accessed from multiple call sites then each one is instrumented separately, so that profiling information is specific to the calling context. %
(6) \emph{Application-specific insight:} the operator can manually disable instrumentation for a map if it is clear from operational context that access patterns prohibit any traffic-dependent optimization (see Table~\ref{tab:optimizations-list}). Traffic-independent optimizations are still applied by \fname in such cases.

\vspaceswrap{0.05in}
\noindent\textbf{Running example.} %
Consider the \texttt{vip\_map} in our sample program, identified as an RO map in the first pass.  In addition, suppose that there are hundreds of VIPs associated with TCP services stored in the \texttt{vip\_map} and only a single one is running QUIC, but the QUIC service receives the vast majority of run-time hits.  Then, instrumentation will identify the QUIC VIP as a heavy hitter and \fname will seize the opportunity to specialize the subsequent QUIC call-path explicitly.  %
Note that this comes without \emph{direct} traffic monitoring, only using \emph{indirect} traffic-specific instrumentation information.

\subsection{Optimization Passes}
\label{sec:dp-optimizations}
The third step of the compilation pipeline is where all online code transformations are applied. Below, we describe the various run-time optimizations; see Table~\ref{tab:optimizations-list} for a summary.

\subsubsection{Just-in-time compilation (JIT)}
\label{sec:jit}
Empirical evidence (see \S\ref{sec:motivations}) suggests that table lookup is a particularly taxing operation for software data planes. This is because certain match-action table types (e.g., LPM or wildcard), that are relatively simple in hardware, are notoriously expensive to implement in software \cite{832493}. Therefore, \fname specializes tables at run time with respect to their content and dynamic access patterns, as learned in the instrumentation pass. Specifically, empty maps are completely removed, small maps are unconditionally just-in-time (JIT) compiled into equivalent code, and larger maps are preceded by a similar JIT compiled fast-path cache, which is in charge of handling the heavy hitters. Note that the consistency of the the fast-path cache must be carefully protected against potential changes made to the specialized map entries; \fname places guards into the code to ensure this (see later).

\vspaceswrap{0.05in}
\noindent\textbf{Running example.} %
Consider again Listing~\ref{lst:katran-ebpf} and suppose that there are only two VIPs configured in the \texttt{vip\_map}. Being an exact-matching hash it is trivial to compile the \texttt{vip\_map} into an ``if-then-else'' statement,  representing each distinct map key as a separate branch. To do so, \fname uses the insights from the code analysis phase to discover that relevant fields in the lookup are the destination address (\texttt{pkt.dstIP}), port (\texttt{pkt.dstPort}) and the IP protocol (\texttt{pkt.proto}). Then, for each entry in the map, it builds a separate ``if'' conditional to compare the entry's fields against the relevant packet header fields and chains these with ``else'' blocks. Since the instrumentation and the just-in-time compiled map are specific to unique combinations of destination address\slash port and protocol, the lookup semantics is correctly preserved even for longest prefix matching (LPM) caches and wildcard lookup.

\subsubsection{Constant propagation}
\label{sec:constant-prop}
Specializing a table does not only benefit the performance of the lookup process: it also has far reaching consequences for the rest of the code. This is because a specialized table contains all the constants (keys and values) \emph{inlined}, which makes it possible to propagate these constants to the surrounding code in order to inline memory accesses. In \fname, constant propagation opportunistically extends to larger maps that cannot be wholly just-in-time compiled: if a certain table field is found to be constant across \emph{all} entries, then this constant is also inlined into the surrounding code. This optimization is thereby two-faceted: it can be used to specialize the code with respect to the inbound traffic (traffic-dependent, former case) but can also be applied without packet-level information (traffic-independent, the latter case). \fname does not implement constant propagation itself; rather, it relies on the underlying compiler toolchain to perform this pass. %

\vspaceswrap{0.05in}
\noindent\textbf{Running example.} %
Suppose there are only two backends in the \texttt{backend\_pool}. Here, the map lookup (line~\ref{line:katran_real_lookup}) is rewritten into an ``if-then-else'' statement, with two branches for each backend. Correspondingly, in each branch the value of the \texttt{backend} variable is constant, which allows to save the costly memory dereference \texttt{backend->ip} (line~\ref{line:katran_encapsulate}) by inlining the backend IP address right into the specialized code.  

\begin{figure*}[t!]
    \begin{subfigure}{.33\textwidth}
        \centering
        \includegraphics[clip,width=0.95\textwidth]{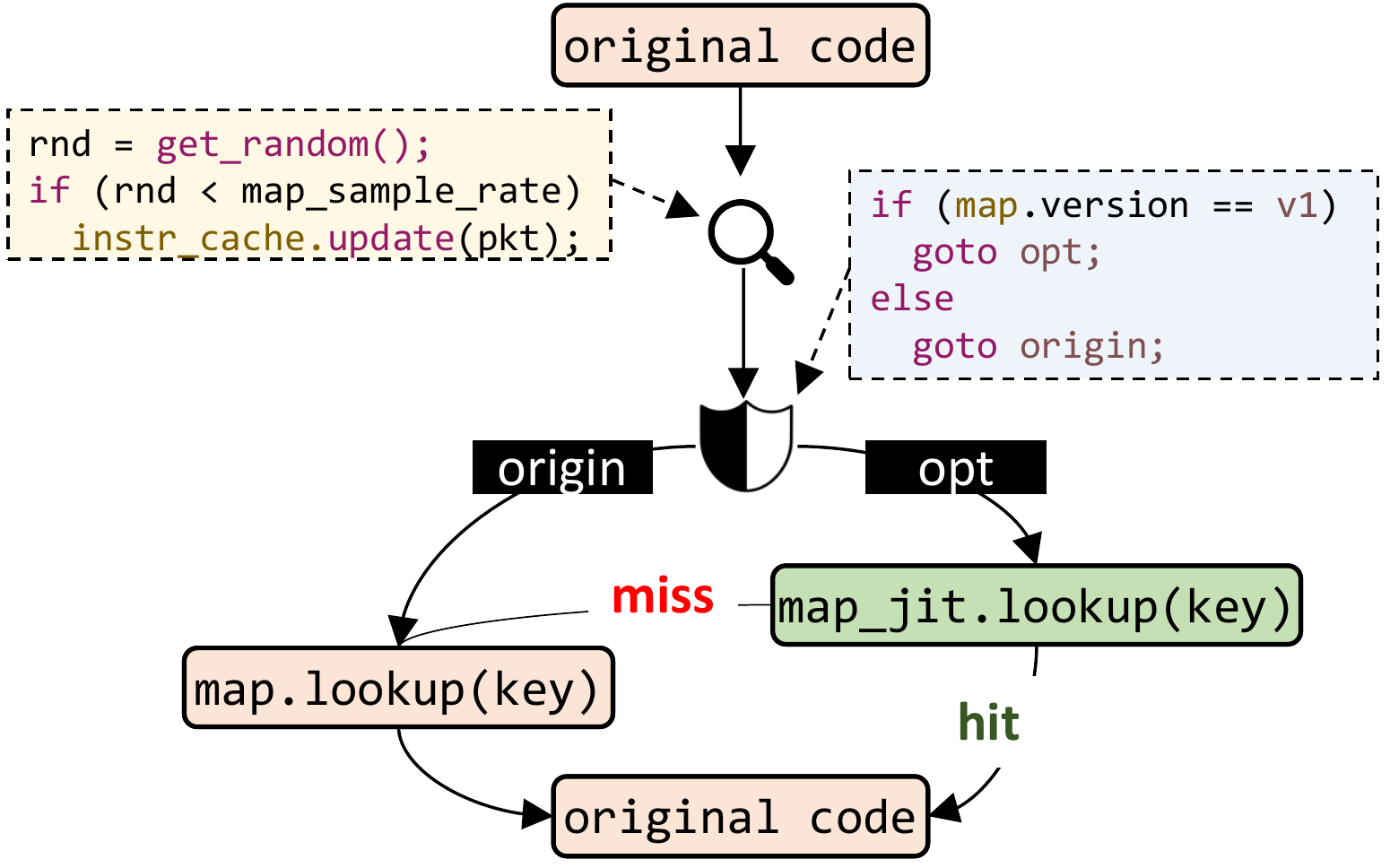}
        \caption{Read-Write table}
        \vspaceswrapfigure{-0.4cm}
        \label{fig:rw-table}
    \end{subfigure}
    \begin{subfigure}{.33\textwidth}
        \centering
        \includegraphics[clip,width=0.75\textwidth]{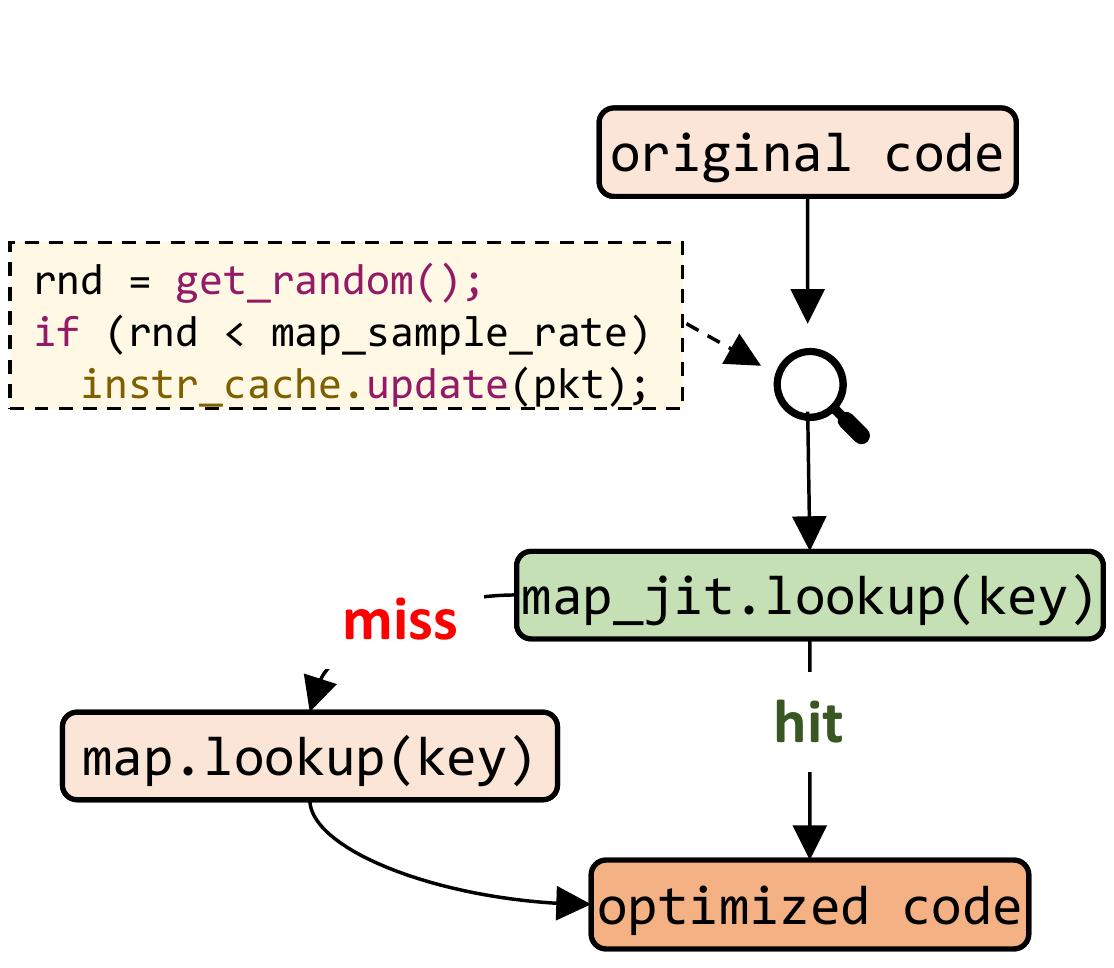}
        \caption{Large Read-Only table}
        \label{fig:large-ro}
    \end{subfigure}
    \begin{subfigure}{.33\textwidth}
        \centering
        \includegraphics[clip,width=0.45\textwidth]{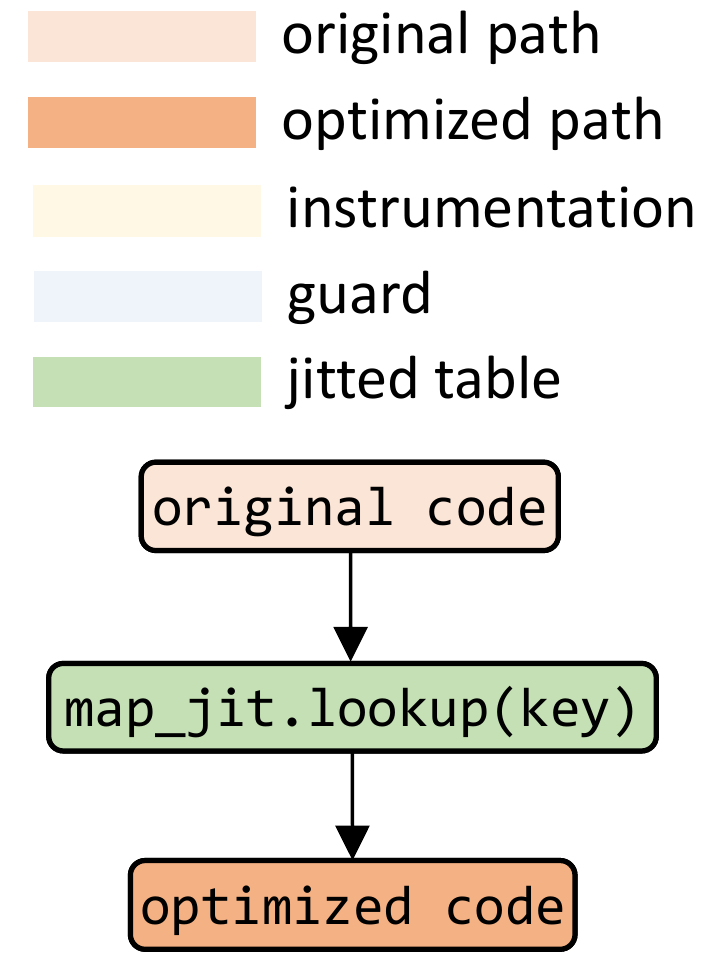}
        \caption{Small Read-Only table}
        \vspaceswrapfigure{-0.2cm}
        \label{fig:small-ro}
    \end{subfigure}
    \caption{\fname handles the optimizations and provide code consistency mechanisms that are table-dependent.} %
    \label{fig:tables-opt}
\end{figure*}

\subsubsection{Dead code elimination}
\label{sec:dce}
Depending on the specific configuration, a large portion of code may sit unused in memory at any point in time. Such ``dead code'' can be found using a combination of static code analysis and the instrumentation information obtained from the previous pass. Upon detection, \fname removes all dead code on the optimized code path. As for the previous case, this operation is outsourced to the underlying compiler. 

\vspaceswrap{0.05in}
\noindent\textbf{Running example.}
Consider the \texttt{vip\_map} lookup site (line~\ref{line:katran_vip_lookup1}) and suppose that there are no QUIC services configured. As a consequence, the \texttt{vip\_info->flags} is identical across all the entries in the \texttt{vip\_map} and the constant propagation pass inlines this constant into the subsequent conditional (line~\ref{line:katran_vip_lookup1}). Thus, the condition \texttt{vip\_info->flags \& F\_QUIC\_VIP} always evaluates to \texttt{false} and the subsequent branch can be safely removed.  %

\subsubsection{Data Structure Specialization}
\label{sec:ds-spec}
\fname adapts the layout, size and lookup algorithm of a table against its content at run time. For example, if all entries share the same prefix length in an LPM map, then  a much faster exact-matching cache \cite{molnar2016dataplane} can be used. This is done by first associating a backend-specific cost function with each applicable representation (this can be automatically inferred using static analysis and symbolic execution \cite{panchenko2019bolt,pedrosa2018automated}), generate then the expected cost of each candidate, and finally implement the table that minimizes the cost.

\subsubsection{Branch Injection}
\label{sec:branch-inj}
This applies to the cases when certain fields take only few possible values in a table. In this situation, it is possible to eliminate subsequent code that handles the \emph{rest} of the values. Such an optimization was used in \S\ref{sec:motivations} to sidestep the ACL lookup for UDP packets in the \emph{firewall}: if we observe that the ``IP protocol'' field can have only a single value in the ACL (e.g., TCP), then we can inject a conditional statement \emph{before} the ACL lookup to check if the IP protocol field in a packet is TCP, use symbolic execution to track the use of this value throughout the resultant branch, and invoke dead code elimination to remove the useless ACL lookup on the non-TCP ``else'' branch.

\subsubsection{Guard elision}
\label{sec:guards}
A critical requirement in dynamic compilation is to protect the consistency of the specialized code against changes to the invariants the optimizations depend on. Such changes can be made from the control plane or, when the program implements a stateful network function, even from the data-plane itself. %
A standard mechanism used by dynamic compilers to guarantee code consistency is to inject simple run-time version checks, called \emph{guards}, at specific points in the code \cite{10.1145/178243.178478}.  When the control flow reaches a guard, it atomically checks if the version of the guard is the same as the version of the optimized code; if yes, execution jumps to the optimized version, otherwise it falls back to the original code (``deoptimization'').  Since each packet may need to pass multiple checks while traversing the datapath, guards may introduce nontrivial run-time overhead \cite{guard-elision}. To mitigate this, \fname heuristically eliminates as many guards as possible; this is achieved by using different schemes to protect stateful and stateless code.

\vspaceswrap{0.05in}
\noindent\textbf{Handling control plane updates.} Theoretically, each table should be protected by a guard when the contents are modified from the control plane. This would require packets to perform one costly guard check for each table. To reduce this overhead, \fname collapses all table-specific guards protecting against control plane updates into a single program-level guard, injected at the program entry point. Once an RO map gets updated by the control plane, the program-level guard direct all incoming packets to the \textit{original} (unoptimized) datapath until the next compilation cycle kicks in to re-optimize the code with respect to the new table content.

\vspaceswrap{0.05in}
\noindent\textbf{Handling updates within the data plane.} The optimized datapath must be protected from data-plane updates as well, which requires an explicit guard at all access sites for RW maps. %
If the guard tests valid then a query is made into the just-in-time compiled fast-path map cache and, on cache hit, the result is used in the subsequent code. Once a modification is made to the map from the program, the guard is invalidated and map lookup falls back to the original map.

Fig.~\ref{fig:tables-opt} presents a breakdown of the strategies \fname uses to protect the consistency of optimized code. For RW maps (Fig.~\ref{fig:rw-table}), first an instrumentation cache is inlined at the access sites, followed by a guard that protects the just-in-time compiled fast-path against data-plane updates. Note that the constant propagation and dead code elimination passes are suppressed, since these passes may modify the code \emph{after} the map lookup and the guard does not protect these optimizations. In contrast, RO map lookups (Fig.~\ref{fig:large-ro} and Fig.~\ref{fig:small-ro}) elide the guard, because only control-plane updates could invalidate the optimizations in this case but these are covered by the program-level guard. RO maps are specialized more aggressively than RW maps, by enabling all optimization passes. Finally, additional overhead can be shaved off for small RO tables by removing the fall-back map all together (Fig.~4c).

\vspaceswrap{0.05in}
\noindent\textbf{Running example.} %
Once static code analysis confirms that the \texttt{vip\_map} and \texttt{backend\_pool} maps are RO, \fname opportunistically eliminates the corresponding guards at the call site. This then implies that, as long as the VIPs and the backend pool are invariant, the optimized code elides the guard.  Since the \texttt{conn\_table} map is RW, it is protected with a specific guard at the call site (line~\ref{line:katran_conn_lkp}).  Thus, the specialized map is used only as long as the connection tracking module's state remains constant; once a new flow is introduced into \texttt{conn\_table} (line~\ref{line:katran_conn_update}) the specialized code is immediately invalidated by bumping the data-plane version. This does not invalidate all optimizations: as long as the rest of the (RO) maps are not updated by the control plane, the program-level guard remains valid and the corresponding RO map specializations still apply.

\subsection{Update}
\label{sec:pipeline-switch}
Upon invocation, \fname executes the above passes to create the optimized datapath and uses the native compiler toolchain to transform the optimized code to target native code.  Meanwhile, control plane updates are temporarily queued without being processed. This allows the ``old'' code to process packets without any disruption while the optimization takes place.  Once compilation is finished, the optimized code is injected into the data path, the program-level guard is updated \cite{10.5555/2772722.2772728} and the outstanding table updates are executed.

\section{Implementation}
\label{sec:implementation}
\fname is implemented in about 5940 lines of C++ code and it is openly available at \url{link-anonymized}.
The code is separated into a data plane independent portable \emph{core}, containing the compiler passes, and technology-specific \emph{plugins} to interact with the underlying technology (i.e., eBPF, DPDK).

The \fname core extends the LLVM~\cite{lattner2004llvm} compiler toolchain (v10.0.1) %
for code manipulation and run-time code generation. %
We opted to implement \fname at the \emph{intermediate representation} (IR) level as it allows to reason about the running code using a relatively high-level language framework without compromising on code generation time. Moreover, this also makes the \fname core portable across different data plane frameworks and programming languages \cite{10.1145/2774993.2775000}.

The data plane plugins are abstracted via a \emph{backend API}. This API exports a set of functions for the core to identify match-action table access sites based on data-plane specific call signatures; compute cost functions for data structure specialization; rewrite data plane dependent code using templates; and provide an interface to inject guards. Additionally, the backend can channel instrumentation data from the data plane to the compiler core, implement the data plane dependent parts of the pipeline update mechanism, and provide a mechanism for the \fname core to intercept, inspect, and queue any update made by the control plane. The latter allows the compilation pipeline to be triggered when \fname intercepts a control plane event, e.g., an update to a table. Currently, only eBPF (fully) and DPDK (partially) are supported, but the architecture is generic enough to be extended to essentially any I/O framework, like netmap \cite{rizzo12} or AF\_XDP \cite{afxdp}.

\subsection{The eBPF Plugin}
\label{sec:polycube-implementation}
\fname leverages the Polycube~\cite{miano2019service} framework as an eBPF backend %
to manage chains of in-kernel packet processing programs. Polycube readily delivers almost all the needed components for an eBPF backend. We added a mechanism for updating the data plane program on-the-fly and defined templates to inject guards. We discuss these components next.

\vspaceswrap{0.1in} %
\noindent \textbf{Pipeline update.} %
Once the optimized program is built, \fname calls the eBPF LLVM backend to generate the final eBPF native code, loads the new program into the kernel using the \texttt{bpf()} system call, and directs execution to the new code. %
In Polycube, a generic data plane program is usually realized as a chain of small eBPF programs connected via the eBPF \textit{tail-call} mechanism, using a \texttt{BPF\_PROG\_ARRAY} map to get the address of the entry point of the next eBPF program to execute. %
Thus, injecting a new version of an eBPF program boils down to atomically update the \texttt{BPF\_PROG\_ARRAY} map entry pointing to it with the address of the new code. %

\vspaceswrap{0.1in} %
\noindent \textbf{Guards.} %
\fname relies on guards to protect the specialized code against map updates. The program-level guard is implemented as a simple run-time version check \cite{10.5555/2772722.2772728}. For stateful processing, \fname installs a guard at each map lookup site and injects a \emph{guard update pre-handler} at the instruction address corresponding to the map update eBPF function (\texttt{map\_update\_elem}). This handler will then safely invalidate the guard before executing the map update. %

\subsection{The DPDK Plugin}
\label{sec:fastclick-implementation}
\fname leverages FastClick~\cite{barbette:fastclick}, a framework to manage packet-processing applications based on DPDK. FastClick makes implementing most components of the backend API trivial; below we report only on pipeline updates and guards.

\vspaceswrap{0.1in}%
\noindent \textbf{Pipeline update.} %
A FastClick program is assembled from primitive network functions, called \textit{elements}, connected into a dataflow graph. Every FastClick element holds a pointer to the next element along the processing chain. To switch between different element implementations at run time, \fname adds a level of \textit{indirection} to the FastClick pipeline: %
every time an element would pass execution to the next one, the corresponding function call is conveyed through a trampoline, which stores the \emph{real} address of the next element to be called. Then, atomic pipeline update simplifies into rewriting the corresponding trampoline to the address of the new code. In contrast to eBPF, which explicitly externalizes into separate maps all program data intended to survive a single packet's context, a FastClick element can hold non-trivial internal \emph{state}, which would need to be tediously copied into the new element. As a workaround, our DPDK plugin disables dynamic optimizations for stateful FastClick elements.

\vspaceswrap{0.1in}%
\noindent \textbf{Guards.} %
Since stateful FastClick elements are never optimized in \fname and RO elements maps always elide the guard, our DPDK plugin currently does not implement guards, except a program-level version check at the entry point.

\section{Evaluation}
\label{sec:evaluation}

Our testbed includes two servers connected back-to-back with a dual-port Intel XL710 40Gbps NIC. The first, a 2x10-core Intel Xeon Silver 4210R CPU @2.40GHz with support for Intel's Data Direct I/O (DDIO)~\cite{intel:ddio} and 27.5 MB of L3 cache, runs the various applications under consideration. The second, a 2x10 Intel Xeon Silver 4114 CPU @2.20GHz with 13.75MB of L3 cache, is used as packet generator.
Both servers are installed with Ubuntu 20.04.2, with the former running kernel 5.10.9 and the latter kernel 4.15.0-112.
We also configured the NIC Receive-Side Scaling (RSS) to redirect all flows to a single receive queue, forcing the applications to be executed on a single CPU core, while \fname was pinned to another CPU core on the device-under-test (DUT). 

In our tests, we used \texttt{pktgen} with DPDK v20.11.0 to generate traffic and report the throughput results.
Unless otherwise stated, we report the average single-core throughput across five different runs of each benchmark, measured at the minimum packet size (64-bytes).
For latency tests, we used \textit{Moongen}~\cite{emmerich2015moongen} to estimate the round-trip-time of a packet from the generator to the DUT and back. Finally, we used \textit{perf} v5.10 to characterize the micro-architectural metrics of the DUT (e.g., cache misses, cycles, number of instructions).

In order to benchmark \fname on real applications, we chose four eBPF/XDP-based packet processing programs from the open-source eBPF/XDP reference network function virtualization framework Polycube~\cite{polycube-tnsm}, plus Facebook's Katran load-balancer used earlier as a running example~\cite{hopps:katran}.

The \emph{L2 switch}, the \emph{Router} and the \emph{NAT} applications were taken from Polycube~\cite{polycube-tnsm}. The \emph{L2 switch} use case is a functional Ethernet switch supporting 802.1Q VLAN and STP, with STP and flooding delegated to the control plane while learning and forwarding implemented entirely in eBPF, using an exact-matching MAC table supporting up to 4K entries. The \emph{Router} use case implements a standard IP router, with RFC-1812 header checks, next-hop processing and checksum rewriting, configured with an LPM table of 2590 prefixes taken from the Stanford routing tables~\cite{kazemian2012stanford}. The \emph{NAT} is an eBPF re-implementation of the corresponding Linux Netfilter application, configured with a single two-way SNAT\slash masquerading rule: the source IP of every packet is replaced with the IP of the outgoing NAT port and a separate L4 source port is allocated for each new flow. \emph{BPF-iptables} is an eBPF/XDP clone \cite{miano2019securing} of the well-known Linux \textit{iptables} framework, configured with 500 wildcard 5-tuple rules generated by Classbench~\cite{taylor2007classbench}. Finally, \emph{Katran} \cite{hopps:katran} was configured as a web-frontend, with 10 TCP services\slash VIPs and 100 backend servers for each VIP. 

For each benchmark, we generated 3 traffic traces with varying locality, to demonstrate the ability of \fname to track packet-level dynamics and optimize the programs accordingly. In particular, we created a \textit{high-locality} traffic trace, where the top-5 flows account for 95\% of the total traffic, a \textit{low-locality} trace where the top-50 flows contribute 95\% of the total traffic, and a \textit{no-locality} trace with \textasciitilde1000 different flows generated at random by a uniform distribution. Classbench comes with built-in trace generator, this was used for the \emph{BPF-iptables} benchmarks. Each flow remains active for the entire duration of the experiments (see later on the dynamic benchmarks).

\subsection{Benefits of Optimizations}
First, we characterize the performance impact of \fname over the mentioned eBPF applications, when attached to the XDP hook of the ingress interface. 

\begin{figure}[t]
	\centering
    \includegraphics[clip=true, width=\linewidth, trim=0cm 0cm 0cm 0cm]{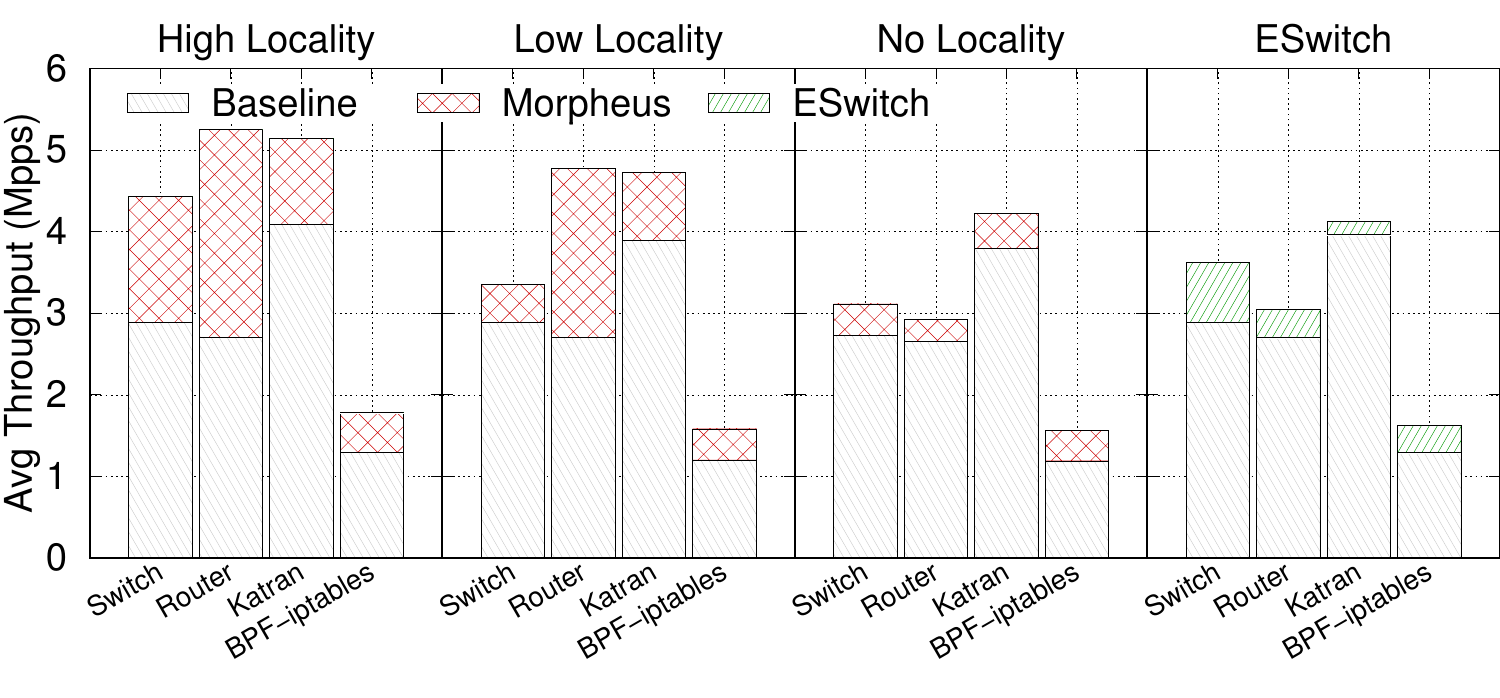}
    \vspaceswrapfigure{-0.3in}
	\caption{Single core throughput (64B packets) varying input traffic locality. The optimizations adopted by \fname are traffic-dependent, while the ones from ESwitch~\cite{molnar2016dataplane} are not.}
	\label{fig:overall-throughput}
\end{figure}

\begin{figure}[t]
	\centering
   	\includegraphics[clip=true, width=1.0\linewidth, trim=0cm 0cm 0cm 0cm]{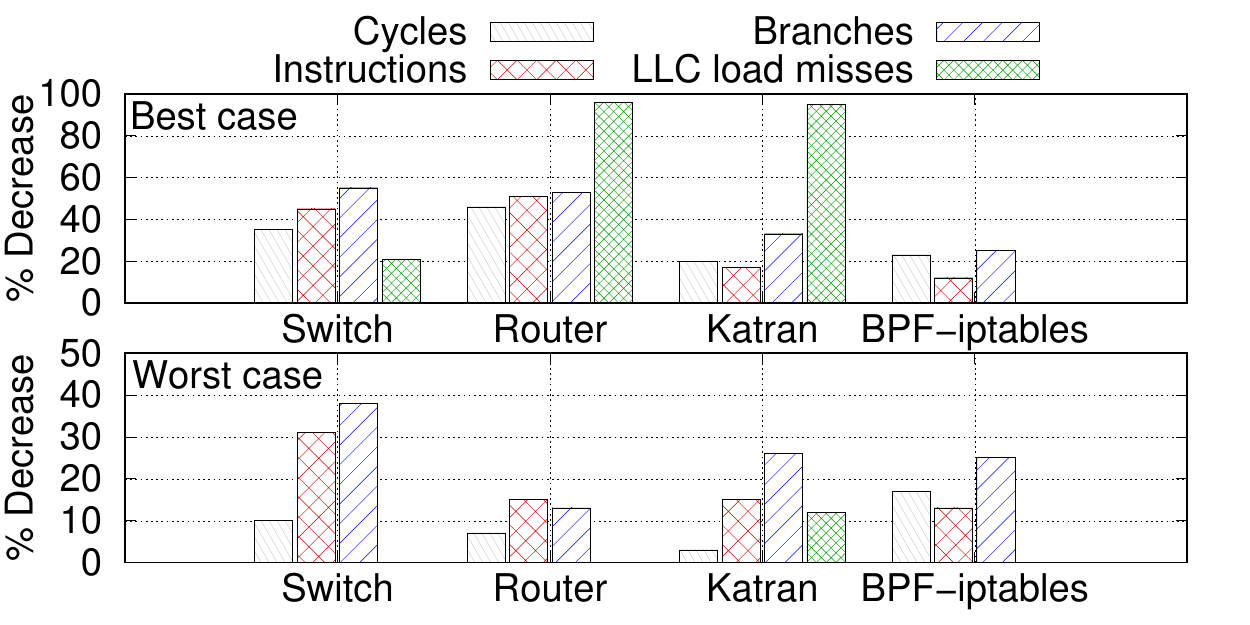}
   	\vspaceswrapfigure{-0.3in}
	\caption{Effect of \fname optimizations on PMU counters, obtained with \textit{perf} at the default frequency (40KHz). The top panel shows the percentage of \textit{decrease}, per packet, of different metrics for \textit{high locality} traffic (best-case for \fname), and the bottom panel for \textit{no locality} traffic (worst-case).}
	\label{fig:overall-perf}
	\vspaceswrapfigure{-0.1in}
\end{figure}

\vspaceswrap{0.05in}
\noindent \textbf{\fname improves packet-processing throughput.}
In Fig.~\ref{fig:overall-throughput}, we show the impact of \fname under different traffic conditions. Throughput is defined as the maximum packet-rate sustained by a program without experiencing packet loss. When a small subset of flows sends the majority of traffic (high-locality), \fname consistently provides more than 50\% throughput improvement over the baseline, with a $2\times$ speed-up for the \emph{Router}. This is because \fname can track heavy flows and optimize the code accordingly. To confirm the benefit of packet-level optimizations in \fname, we compared it to a faithful eBPF/XDP re-implementation of ESwitch, a dynamic compiler that does not consider traffic dynamics~\cite{molnar2016dataplane}. The results (Fig~\ref{fig:overall-throughput}) clearly show that \fname consistently delivers $5$--$10\times$ the improvement compared to ESwitch for high-locality traces, while it essentially falls back to ESwitch for uniform traffic.

\vspaceswrap{0.05in}
\noindent \textbf{\fname benefits at the micro-architectural scale.}
Fig.~\ref{fig:overall-perf} confirms that, by specializing code for the input the program is processing, it allows packet-processing programs to execute more efficiently. \fname reduces the last-level CPU cache misses by up to 96\% and effectively halves the instructions and branches executed per packet. At low or no traffic locality, the effects of packet-level optimizations diminish, but \fname can still bring considerable performance improvement; e.g., we see $\sim 30\%$ margin for \emph{BPF-iptables} even for the \emph{no-locality} trace. This is because the optimization passes in \fname are carefully selected to be applicable independently from packet-level dynamics (see Table~\ref{tab:optimizations-list}). %

\begin{figure}[t]
	\centering
    \includegraphics[clip=true, width=.96\linewidth, trim=0cm 0cm 0cm 0cm]{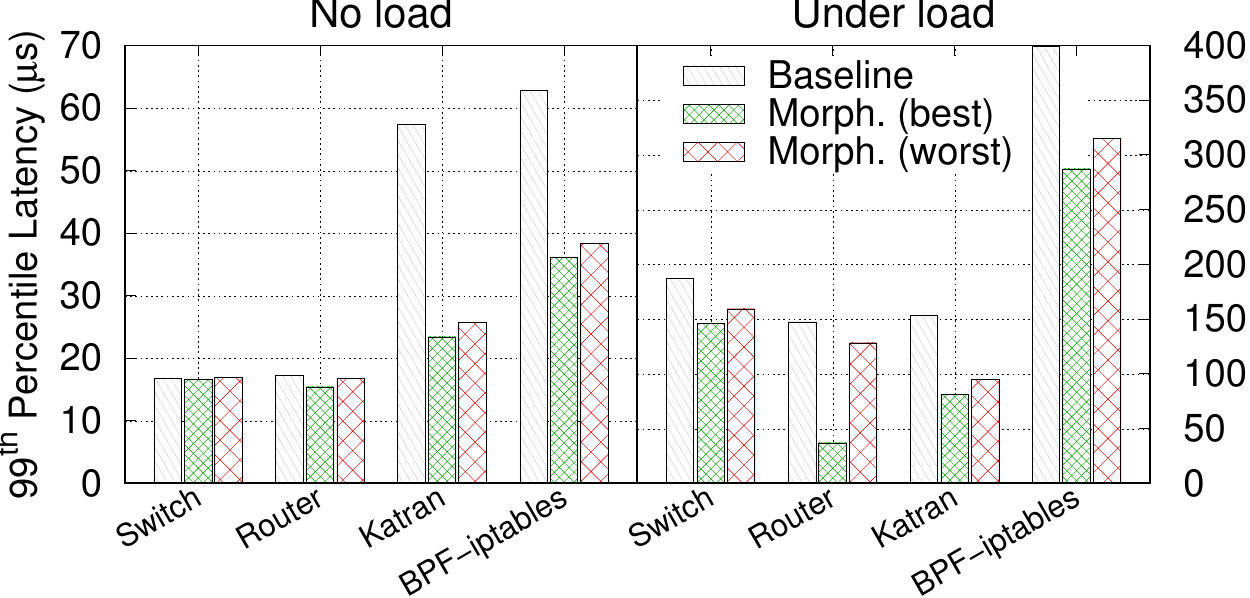}
    \vspaceswrapfigure{-0.1in}
	\caption{99th percentile latency with \fname. The graph shows both the latency for the \textit{optimized} and \textit{non-optimized} code paths, under small load (10pps) and heavy load (highest rate without packet drop).} %
	\label{fig:overall-latency}
	\vspaceswrapfigure{-0.2in}
\end{figure}

\vspaceswrap{0.05in}
\noindent \textbf{\fname reduces packet-processing latency.}
We compared the 99th percentile baseline latency for each application against the one obtained with \fname, both in a \emph{best-case scenario} when all packets travel through the optimized code path (e.g., the right branch in Fig.~\ref{fig:rw-table}), and a \emph{worst-case scenario} with all packets falling back to the default branch instead of taking the fast-patch cache for each map (the left branch in Fig.~\ref{fig:rw-table}). %
The left panel in Fig.~\ref{fig:overall-latency} shows the latency measured at low packet rate (10pps) so to avoid queuing effects~\cite{rfc1242}, whereas the right panel shows latency under the maximum sustained load without packet drops~\cite{rfc2544}. %
First, we observe that \fname never \emph{increases} latency, despite the considerable additional logic it injects dynamically into the code (guards, instrumentation; see below); in fact, it generally reduces it even in the worst case scenario. %
Notably, it reduces Katran's packet-processing latency by about 123\%.

\subsection{What is the cost of code instrumentation?}
\label{sec:instrumentation-overhead}
\begin{figure}[t]
	\centering
	\includegraphics[clip=true, width=1.0\linewidth, trim=0cm 0cm 0cm 0cm]{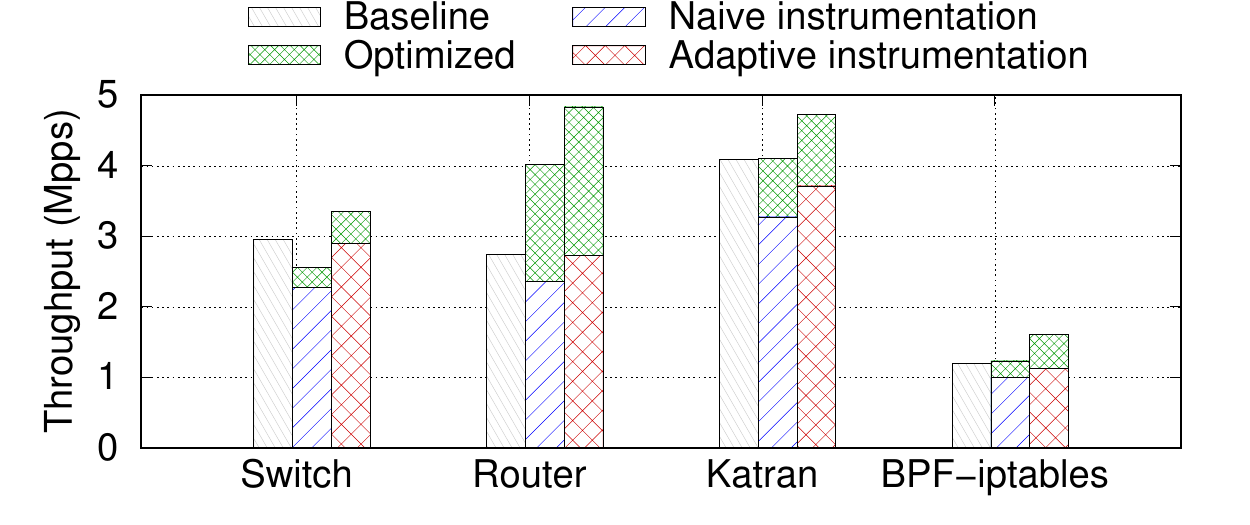}
	\vspaceswrapfigure{-0.3in}
	\caption{Naive vs adaptive instrumentation (low locality traffic). In the naive case all map lookups are recorded, while adaptive instrumentation adjusts data sampling selectively for the access patterns at each lookup call site.}
	\label{fig:naitve-vs-adaptive-instrumentation}
	\vspaceswrapfigure{-0.1in}
\end{figure}

Clearly, the price for performance improvements is the additional logic, most prominently, instrumentation, injected by \fname into the fast packet-processing path.  
To understand this price, we compared our \textit{adaptive} instrumentation scheme (\S\ref{sec:runtime-stats-collection}) against a \textit{naive} approach where all map lookups are explicitly recorded. %
Fig.~\ref{fig:naitve-vs-adaptive-instrumentation} shows that instrumentation involves visible overhead: the instrumented code performs worse than the baseline. The naive approach imposes a hefty 14--23\% overhead, but adaptive instrumentation reduces this to just 0.9\%--9\%. Most importantly, this reduction does not come at a prohibitive cost: adaptive instrumentation provides enough insight to \fname to make up for the performance penalty imposed by it and still attain a considerable throughput improvement on top (see the green stacked barplots). 
In contrast, the performance tax of naive instrumentation may very well nullify optimization benefits, even despite full visibility into run-time dynamics (e.g., for the L2 switch or Katran).

\begin{figure}[t]
	\centering
	\includegraphics[clip=true, width=1.0\linewidth, trim=0cm 0cm 0cm 0cm]{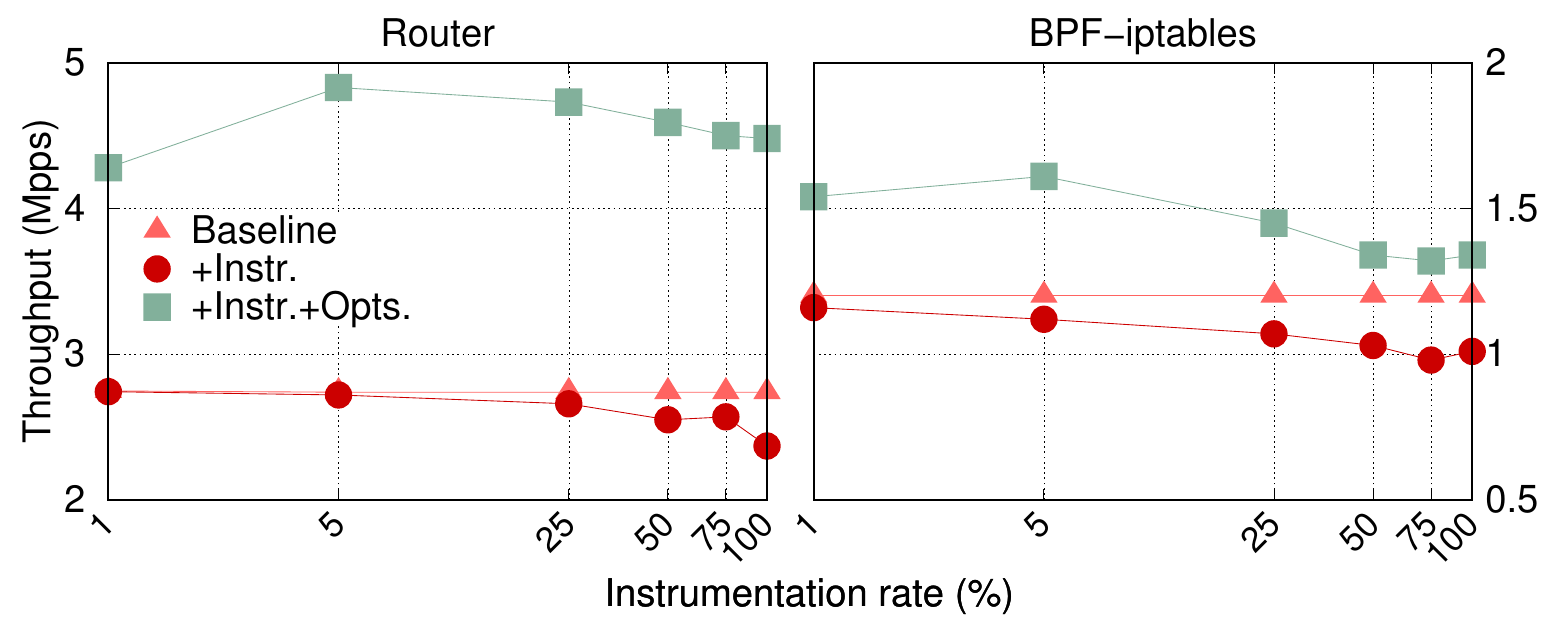}
	\vspaceswrapfigure{-0.3in}
	\caption{Effectiveness of instrumentation at varying sampling rates (Router and BPF-iptables, low-locality traffic).}
	\label{fig:instr-overhead-router-firewall}
	\vspaceswrapfigure{-0.1in}
\end{figure}

We also studied the impact of packet sampling rate on instrumentation. Indeed, \fname collects information on packet-level dynamics only on a subset of input traffic in order to minimize the overhead. Fig.~\ref{fig:instr-overhead-router-firewall} highlights that \fname can strike a balance between overhead and efficiency by adapting the sampling rate. At a low sampling rate (e.g., recording every 100th packet) \fname does not have enough visibility into dynamics, which renders traffic-dependent optimizations less effective (but the traffic-invariant optimizations still apply). Higher sampling rates provide better visibility but also impose higher overhead. At the extreme (\emph{BPF-iptables}, 100\% instrumentation rate), optimization is just enough to offset the price of instrumentation. 
In conclusion, we found that setting the sampling rate at 5\%--25\% represents the best compromise. %

\subsection{How fast is the compilation?}
In Table~\ref{tab:compilation-time}, we indicate with t\textsubscript{1} the time to analyze, instrument and optimize the LLVM IR code, 
and with t\textsubscript{2} the time it takes to generate the final eBPF code, starting from the LLVM IR. Note that t\textsubscript{1} is highly dependent on table size: the bigger the tables, the more time needed to read and analyze them. We show the results for high-locality traffic, which we consider the \textit{best case} since \fname needs to track fewer flows, thus requiring lighter instrumentation tables that are faster to analyze, and a \emph{worst case} when traffic with no-locality is fed into the program.
In general, table read time (i.e., t\textsubscript{1}) dominates the compilation time, consistently staying below $100$ms and reaching only for Katran in the worst-case scenario almost $600$ms. This is because Katran uses huge static maps containing tens of thousands of entries to implement consistent hashing. Recent advances in the Linux kernel allow to read maps in batches, which would cut down this time by as much as 80\% \cite{bpf_map_batch}, reducing recompilation time for Katran below 100ms. Finally, the time needed to inject the optimized datapath into the kernel depends on the complexity of the program, since all eBPF code must pass the in-kernel verifier for a safety check before being activated. This also ensures that a mistaken \fname optimization pass will never break the data plane. In our tests, injection time varies between 0.5 to 3.4ms in the best case and at most 6.1ms in the worst case.

\begin{savenotes}
\begin{table}[t]
\scriptsize
\caption{Time (in ms) to execute the entire \fname compilation pipeline and install the optimized datapath. LOC is calculated using \textit{cloc} (v1.82) excluding comments and blank lines while instruction count is measured with \textit{bpftool} v5.9.}
\centering
\renewcommand{\arraystretch}{1.3}
\setlength{\tabcolsep}{5.5pt}
\begin{threeparttable}
\vspaceswrapfigure{-0.1in}
\begin{tabular}{|l|c|c|c|c|c|c|c|c|}
\hline
\multirow{3}{*}{\textbf{Application}} & \multirow{3}{*}{\textbf{\begin{tabular}[c]{@{}c@{}}C \\ LOC\end{tabular}}} & \multirow{3}{*}{\textbf{\begin{tabular}[c]{@{}c@{}}BPF \\ Insn\end{tabular}}} & \multicolumn{4}{c|}{\textbf{Compilation (ms)}}    & \multicolumn{2}{c|}{\textbf{Injection (ms)}} \\ \cline{4-9} 
& &  & \multicolumn{2}{c|}{\textbf{Best}} & \multicolumn{2}{c|}{\textbf{Worst}} & \multirow{2}{*}{\textbf{Best}} & \multirow{2}{*}{\textbf{Worst}} \\ \cline{4-7}
& & & \textbf{t\textsubscript{1}} & \textbf{t\textsubscript{2}} & \textbf{t\textsubscript{1}} & \textbf{t\textsubscript{2}} &                                & \\ \hline
L2 Switch 	 & 243  & 464  & 81 & 62 & 140 & 78 & 0.5 & 0.9 \\ \hline
Router       & 331  & 458  & 87 & 65 & 196 & 91 & 1.1 & 1.3 \\ \hline
BPF-iptables\tnote{*} & 220 & 358 & 95 & 62 & 105 & 87 & 0.6 & 0.5 \\ \hline
Katran       & 494 & 905  & 287 & 115 & 569 & 151 & 3.4 & 6.1 \\ \hline
\end{tabular}
\begin{tablenotes}
\item[*] Uses a chain of eBPF programs; since \fname optimizes every eBPF program separately, values shown refer to the most complex program in the chain.
\item[\textbf{t\textsubscript{1}}] Time to analyze the program, instrument it and read the maps.
\item[\textbf{t\textsubscript{2}}] Time to generate the final eBPF code.
\end{tablenotes}
\end{threeparttable}
\label{tab:compilation-time}
\vspaceswrapfigure{-0.2in}
\end{table}
\end{savenotes}

\subsection{\fname in action}
\begin{figure}[t]
	\centering
	\includegraphics[clip=true, width=1.0\linewidth, trim=0cm 0cm 0cm 0cm]{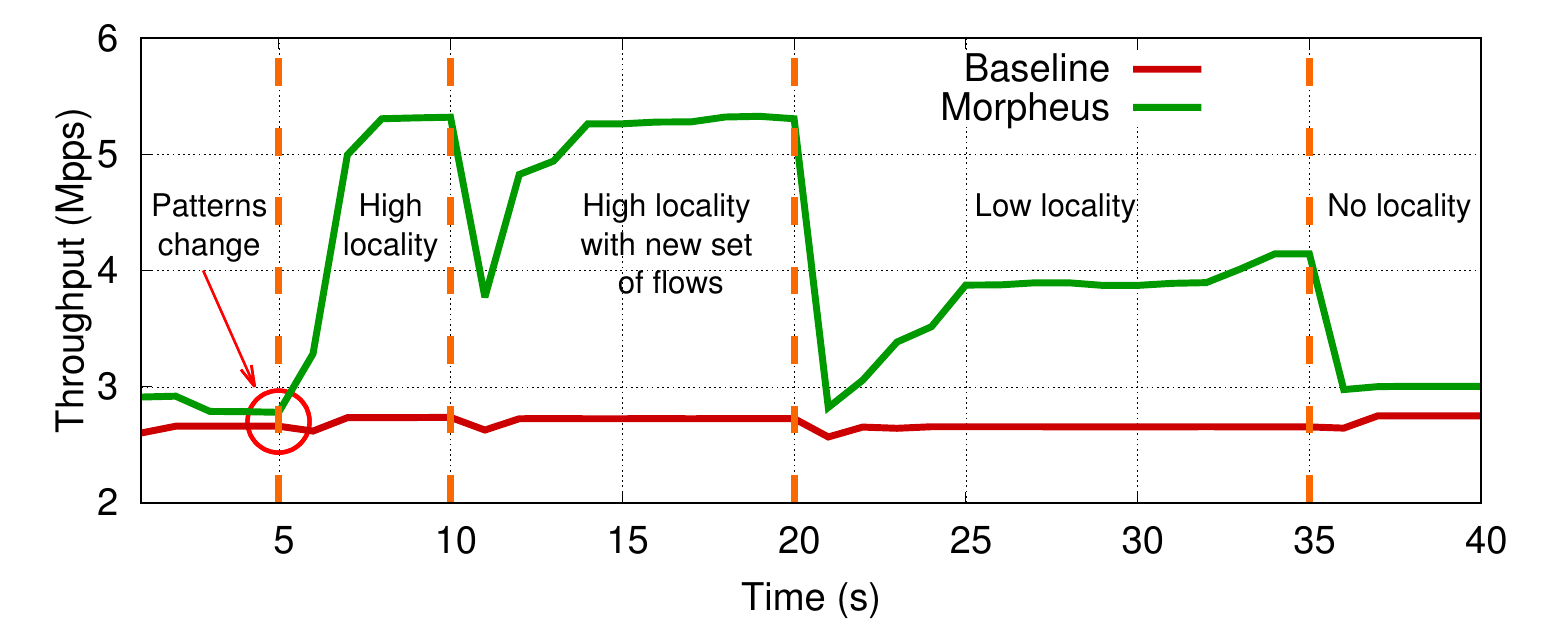}
	\vspaceswrapfigure{-0.2in}
    \caption{Throughput over time with \fname on the \emph{Router} use case, with dynamically changing traffic patterns.}
	\label{fig:router-dynamic-traffic}
	\vspaceswrapfigure{-0.2in}
\end{figure}

To test the ability of \fname to track highly dynamic inputs, we fed the \emph{Router} application with time-varying traffic and observed the throughput over the time (Fig.~\ref{fig:router-dynamic-traffic}).  Recompilation period was conservatively set to 1 second. In the first 5 seconds we generate uniform traffic; here, the traffic-independent optimizations applied by \fname yield roughly 15\% performance improvement over the baseline. At the 5th second, the traffic changes to a high-locality profile: after a quick learning period \fname specializes the code, essentially doubling the throughput. We see the same effect from the 10th second, when we switch to another high-locality trace with a new set of heavy-hitters, and also at 20 seconds, when we switch to a low-locality profile: after a brief training period \fname dynamically adapts the optimized datapath to the new profile and attains 60--100\% performance improvement.

\subsection{What can go wrong?}
\label{sec:nat}
The flip side of dynamic optimization is the potential for a misguided run-time code transformation to harm performance. With generic languages this can happen when the dynamic compiler steals CPU cycles from the running code \cite{591653, stack-java}; in such cases careful manual compiler parameter tuning and deep application-specific knowledge is needed to make up for the lost performance \cite{java-tweak}. Similar issues may arise with dynamically optimizing \emph{network code}, as we show below on the \emph{NAT} use case~\cite{polycube-tnsm}. The NAT is organized as a single large connection tracking table, updated from within the data plane on each new flow. This represents a worst-case scenario for \fname: fully stateful code, so that guards cannot be opportunistically elided, coupled with potentially very high traffic dynamics beyond our control. Yet, since traffic-independent optimizations can still be applied (Table~\ref{tab:optimizations-list}) \fname can improve throughput by around 5\% (from 4.36 to 4.58 Mpps) in the presence of high-locality traffic. However, for low-locality traffic we see about 6\% performance degradation compared to the baseline. Intuitively, \fname just keeps on recompiling the conntrack fast-path with another set of potential heavy hitters, just to immediately remove this optimization as a new flow arrives. Our tests again mark micro-architectural reasons behind this: the number of branch misses and instruction cache loads increases by 90\% and 75\%, respectively, both clear symptoms of frequent code changes. The rest of the stateful applications (\emph{L2 switch} and \emph{Katran}) exhibit a similar pattern, but the speed-up enabled by dead code elimination, constant propagation and branch-injection can make up for this. As with Java, such cases require human intervention; manually disabling optimization for the connection tracking module's table safely eliminates the performance degradation on the NAT use case.

\subsection{\fname with DPDK programs}
\label{sec:eval-morph-dpdk}

\begin{figure}[t]
	\centering
	\includegraphics[clip=true, width=1.0\linewidth, trim=0cm 0cm 0cm 0cm]{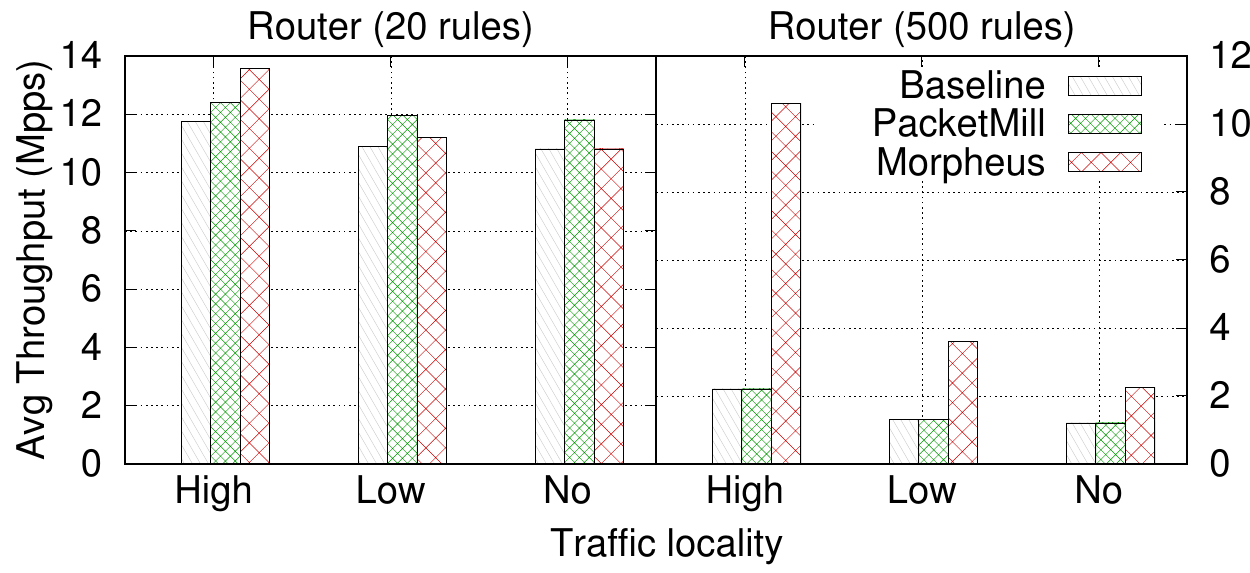}
	\vspaceswrapfigure{-0.2in}
	\caption{Comparison between vanilla FastClick, PacketMill and Morpheus for the Router FastClick application with 20 and 500 rules.}
	\label{fig:eval-throughput-dpdk}
	\vspaceswrapfigure{-0.1in}
\end{figure}

We also applied \fname to a DPDK program, namely the FastClick~\cite{barbette:fastclick} version of the eBPF \emph{Router} application. We configured the router with either 20 or 500 rules taken from the Stanford routing tables~\cite{kazemian2012stanford} and generated traffic with different levels of locality as before. We tested the throughput and the latency of the baseline code and the \fname optimized one and we compared the results to a state-of-the-art DPDK packet-processing optimizer, PacketMill~\cite{packetmill}. In our tests PacketMill uses the following optimizations: removing virtual function calls, inlining variables, and allocating/defining the elements’ objects in the source code.

\begin{figure}[t]
	\centering
	\includegraphics[clip=true, width=1.0\linewidth, trim=0cm 0cm 0cm 0cm]{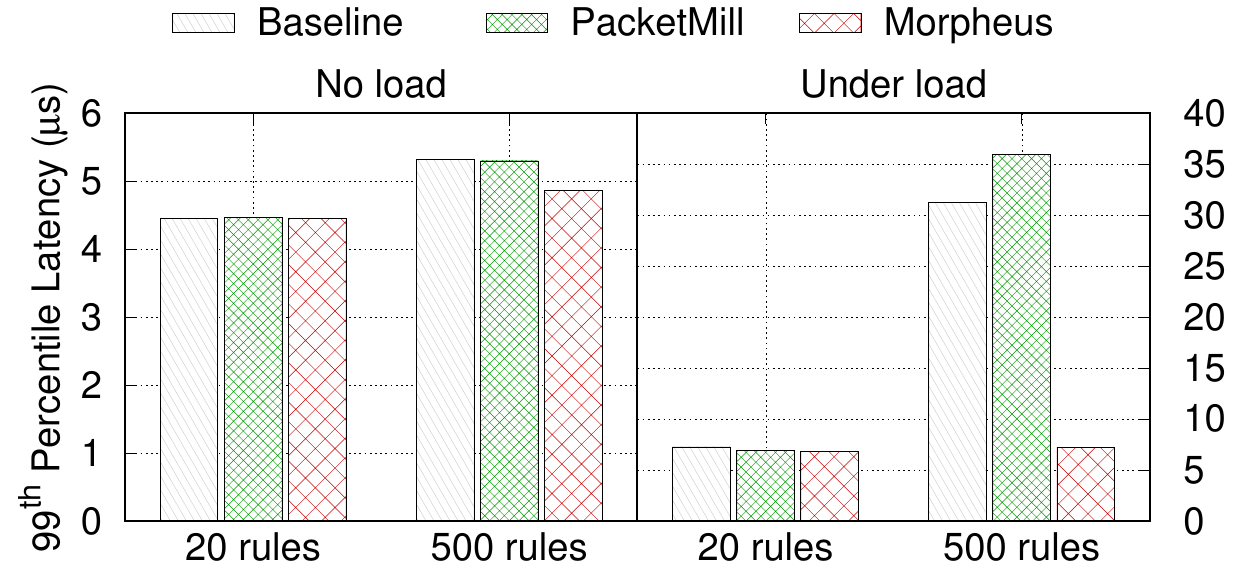}
	\caption{Comparison between vanilla FastClick, PacketMill and Morpheus for the router FastClick application with 20 and 500 rules.}
	\label{fig:eval-latency-dpdk}
	\vspaceswrapfigure{-0.2in}
\end{figure}

Fig.~\ref{fig:eval-throughput-dpdk} reports the average throughput results. For only 20 prefix rules and with low locality traffic, PacketMill outperforms \fname by about 9\%, whereas for high-locality traffic and larger forwarding tables \fname produces a whopping 469\% improvement over PacketMill. The reason for the large performance drop from 20 rules to 500 rules is that LPM lookup is particularly expensive in FastClick (linear search), but \fname can largely avoid this costly lookup by inlining heavy hitters. The 99th percentile latency results (Fig.~\ref{fig:eval-latency-dpdk}) confirm this finding, with \fname decreasing latency 5-fold compared to PacketMill with high-locality traffic. 

\section{Related work}
\label{sec:related-works}

Generic code optimization has a long-standing stream of research and prototypes~\cite{souper, bansal2006automatic, joshi2002denali, 10.1145/3428245, 10.1145/2872362.2872387,autofdo,panchenko2019bolt,propellerllvm}. In the context of networking, domain-specific data-plane optimization has also gained substantial interest lately. %

\vspaceswrap{0.05in}
\noindent
\textbf{Static optimization of data-plane programs.}
Several packet I/O frameworks present specific APIs for developers to optimize network code~\cite{han2015softnic,choi2017pvpp,choi2017case,panda2016netbricks,vpp}, or implement different paradigms to efficiently execute packet-processing programs sequentially or in parallel~\cite{metron,liu2018microboxes,martins2014clickos,ballani2015enabling,sekar2012design,10.1145/3098822.3098826}. Other proposals aim to remove redundant logic or merge different elements together~\cite{openbox, 180672, katsikas2016snf}.
These works, however, provide predominantly \emph{static} optimizations; \fname, on top of these static optimizations, also considers run-time insight to specialize generic network code.

\vspaceswrap{0.05in}
\noindent
\textbf{Dynamic optimization of packet-processing programs.}
ESwitch~\cite{molnar2016dataplane, retvari2017dynamic} was the first functional framework for the unsupervised \emph{dynamic} optimization of software data planes with respect to the packet-processing program, specified in OpenFlow, being executed. PacketMill~\cite{packetmill} and NFReducer~\cite{nfreducer} leverage the LLVM toolchain~\cite{lattner2004llvm} instead of OpenFlow: PacketMill targets the FastClick datapath by exploiting the DPDK packet I/O framework and NFReducer aims to eliminate redundant logic from generic packet-processing programs using symbolic execution.  \fname is strictly complementary to these works: (1) it applies some of the same optimizations but it also introduces a toolbox of new ones (e.g., branch injection or constant propagation for stable table entries); (2) \fname can detect packet-level dynamics and apply more aggressive optimizations depending on the specific traffic patterns; and (3) \fname is data-plane agnostic, in that it performs the optimizations at the IR-level using a portable compiler core and relies on the built-in compiler toolchain to generate machine code and a data-plane plugin to inject it into the datapath.

\vspaceswrap{0.05in}
\noindent
\textbf{Profile-guided optimization for packet-processing hardware.} %
P2GO~\cite{wintermeyer20} and P5~\cite{abhashkumar2017p5} apply several profile-driven optimizations to improve the resource utilization of programmable P4 hardware targets. Some of the ideas presented in this work can also be used with programmable P4 hardware, provided it is possible to re-synthesize the packet processing pipeline without traffic disruption, with a notable difference: P2GO and P5 require \emph{a priori} knowledge (i.e., the profiles) while \fname aims at \emph{unsupervised} dynamic optimization.

\section{Conclusion}
\label{sec:conclusion}
We presented \fname, a run-time compiler and optimizer framework for arbitrary networking code. We demonstrated the importance of tracking packet-level dynamics and how they open up opportunities for a number of domain-specific optimizations. We proposed a solution, \fname, capable of applying them without any \emph{a priori} information on the running program and implemented on top of the LLVM JIT compiler toolchain at the IR level. This allows to decouple  our system from the specific framework used by the underlying data plane as much as possible. Finally, we demonstrated the effectiveness of \fname on a number of programs written in eBPF and DPDK and released the code in open-source to foster reproducibility of our results.

We consider \fname only as a first step towards more intelligent systems that can adapt to network conditions.
As future work, we intend to integrate a run-time performance prediction model~\cite{manousis2020contention, iyer2019performance, pedrosa2018automated, bhardwaj2017preliminary, rath2017symperf} into \fname, which enables the compiler to reason about the effect of each different dynamic optimization pass. This would allow for selecting the most efficient subset of optimizations and adapt the recompilation timescales to the current network conditions.

\def\UrlBreaks{\do\/\do-}
\bibliographystyle{plain}
\bibliography{reference}

\begin{thebibliography}{10}

\bibitem{intel:ddio}
{Intel Data Direct I/O Technology}, Feb 2021.
\newblock
  \url{https://www.intel.co.uk/content/www/uk/en/io/data-direct-i-o-technology.html}.

\bibitem{afxdp}
{Linux AF\_XDP}, Feb 2021.
\newblock \url{https://www.kernel.org/doc/html/latest/networking/af_xdp.html}.

\bibitem{aliasan}
{LLVM Alias Analysis}, Feb 2021.
\newblock \url{https://llvm.org/docs/AliasAnalysis.html}.

\bibitem{memoryssa}
{LLVM MemorySSA}, Feb 2021.
\newblock \url{https://llvm.org/docs/MemorySSA.html}.

\bibitem{abhashkumar2017p5}
Anubhavnidhi Abhashkumar, Jeongkeun Lee, Jean Tourrilhes, Sujata Banerjee,
  Wenfei Wu, Joon-Myung Kang, and Aditya Akella.
\newblock {P5}: Policy-driven optimization of {P4} pipeline.
\newblock In {\em Proceedings of the Symposium on SDN Research}, SOSR ’17,
  page 136–142, New York, NY, USA, 2017. Association for Computing Machinery.

\bibitem{alipourfard2018decoupling}
Omid Alipourfard and Minlan Yu.
\newblock Decoupling algorithms and optimizations in network functions.
\newblock In {\em Proceedings of the 17th ACM Workshop on Hot Topics in
  Networks}, pages 71--77, 2018.

\bibitem{10.1145/231379.231409}
Joel Auslander, Matthai Philipose, Craig Chambers, Susan~J. Eggers, and
  Brian~N. Bershad.
\newblock Fast, effective dynamic compilation.
\newblock In {\em Proceedings of the ACM SIGPLAN 1996 Conference on Programming
  Language Design and Implementation}, PLDI '96, page 149–159, New York, NY,
  USA, 1996. Association for Computing Machinery.

\bibitem{cilium:ebpf-docs}
Cilium Authors.
\newblock Bpf and xdp reference guide.
\newblock feb 2019.
\newblock \url{https://cilium.readthedocs.io/en/latest/bpf/}.

\bibitem{istio}
Istio Authors.
\newblock {Istio - Connect, secure, control, and observe services}, nov 2020.

\bibitem{openstack}
OpenStack Authors.
\newblock Openstack, oct 2020.

\bibitem{10.1145/358438.349303}
Vasanth Bala, Evelyn Duesterwald, and Sanjeev Banerjia.
\newblock {Dynamo}: A transparent dynamic optimization system.
\newblock {\em SIGPLAN Not.}, 35(5):1–12, May 2000.

\bibitem{ballani2015enabling}
Hitesh Ballani, Paolo Costa, Christos Gkantsidis, Matthew~P Grosvenor, Thomas
  Karagiannis, Lazaros Koromilas, and Greg O'Shea.
\newblock Enabling end-host network functions.
\newblock {\em ACM SIGCOMM Computer Communication Review}, 45(4):493--507,
  2015.

\bibitem{bansal2006automatic}
Sorav Bansal and Alex Aiken.
\newblock Automatic generation of peephole superoptimizers.
\newblock {\em ACM SIGARCH Computer Architecture News}, 34(5):394--403, 2006.

\bibitem{8535089}
D.~{Barach}, L.~{Linguaglossa}, D.~{Marion}, P.~{Pfister}, S.~{Pontarelli}, and
  D.~{Rossi}.
\newblock High-speed software data plane via vectorized packet processing.
\newblock {\em IEEE Communications Magazine}, 56(12):97--103, December 2018.

\bibitem{barbette:fastclick}
T.~{Barbette}, C.~{Soldani}, and L.~{Mathy}.
\newblock Fast userspace packet processing.
\newblock In {\em 2015 ACM/IEEE Symposium on Architectures for Networking and
  Communications Systems (ANCS)}, pages 5--16, 2015.

\bibitem{barbette2020ceetah}
Tom Barbette, Chen Tang, Haoran Yao, Dejan Kosti{\'c}, Gerald Q.~Maguire Jr.,
  Panagiotis Papadimitratos, and Marco Chiesa.
\newblock A high-speed load-balancer design with guaranteed
  per-connection-consistency.
\newblock In {\em 17th {USENIX} Symposium on Networked Systems Design and
  Implementation ({NSDI} 20)}, pages 667--683, Santa Clara, CA, February 2020.
  {USENIX} Association.

\bibitem{bhardwaj2017preliminary}
Ankit Bhardwaj, Atul Shree, V~Bhargav Reddy, and Sorav Bansal.
\newblock A preliminary performance model for optimizing software packet
  processing pipelines.
\newblock In {\em Proceedings of the 8th Asia-Pacific Workshop on Systems},
  page~26. ACM, 2017.

\bibitem{rfc1242}
Scott Bradner.
\newblock {Benchmarking Terminology for Network Interconnection Devices}.
\newblock {RFC} 1242, {RFC Editor}, July 1991.

\bibitem{rfc2544}
Scott Bradner and Jim McQuaid.
\newblock Benchmarking methodology for network interconnect devices.
\newblock RFC 2544, RFC Editor, March 1999.
\newblock \url{http://www.rfc-editor.org/rfc/rfc2544.txt}.

\bibitem{openbox}
Anat Bremler-Barr, Yotam Harchol, and David Hay.
\newblock Openbox: A software-defined framework for developing, deploying, and
  managing network functions.
\newblock In {\em Proceedings of the 2016 ACM SIGCOMM Conference}, SIGCOMM '16,
  page 511–524, New York, NY, USA, 2016. Association for Computing Machinery.

\bibitem{autofdo}
Dehao Chen, David~Xinliang Li, and Tipp Moseley.
\newblock Autofdo: Automatic feedback-directed optimization for warehouse-scale
  applications.
\newblock In {\em Proceedings of the 2016 International Symposium on Code
  Generation and Optimization}, CGO ’16, page 12–23, New York, NY, USA,
  2016. Association for Computing Machinery.
\newblock \url{https://doi.org/10.1145/2854038.2854044}.

\bibitem{choi2017case}
Sean Choi, Xiang Long, Muhammad Shahbaz, Skip Booth, Andy Keep, John Marshall,
  and Changhoon Kim.
\newblock The case for a flexible low-level backend for software data planes.
\newblock In {\em Proceedings of the First Asia-Pacific Workshop on
  Networking}, pages 71--77. ACM, 2017.

\bibitem{choi2017pvpp}
Sean Choi, Xiang Long, Muhammad Shahbaz, Skip Booth, Andy Keep, John Marshall,
  and Changhoon Kim.
\newblock Pvpp: A programmable vector packet processor.
\newblock In {\em Proceedings of the Symposium on SDN Research}, pages
  197--198. ACM, 2017.

\bibitem{591653}
T.~{Cramer}, R.~{Friedman}, T.~{Miller}, D.~{Seberger}, R.~{Wilson}, and
  M.~{Wolczko}.
\newblock Compiling {Java} just in time.
\newblock {\em IEEE Micro}, 17(3):36--43, 1997.

\bibitem{nfreducer}
Bangwen Deng, Wenfei Wu, and Linhai Song.
\newblock Redundant logic elimination in network functions.
\newblock In {\em Proceedings of the Symposium on SDN Research}, SOSR '20, page
  34–40, New York, NY, USA, 2020. Association for Computing Machinery.

\bibitem{dpdk:pktgen}
DPDK.
\newblock Pktgen traffic generator using dpdk, aug 2018.

\bibitem{dpdk:l3fwd-acl}
DPDK.
\newblock L3 forwarding with access control sample application, 2021.

\bibitem{emmerich2015moongen}
Paul Emmerich, Sebastian Gallenm\"{u}ller, Daniel Raumer, Florian Wohlfart, and
  Georg Carle.
\newblock Moongen: A scriptable high-speed packet generator.
\newblock IMC '15, page 275–287, New York, NY, USA, 2015. Association for
  Computing Machinery.

\bibitem{10.1145/510726.510749}
Cristian Estan and George Varghese.
\newblock New directions in traffic measurement and accounting.
\newblock {\em SIGCOMM Comput. Commun. Rev.}, 32(1):75, January 2002.

\bibitem{packetmill}
A.~Farshin, T.~Barbette, Roozbeh A, G.~Maguire, and Dejan Kosti'c.
\newblock {PacketMill}: Toward per-core {100-Gbps} networking.
\newblock ASPLOS, 2021.

\bibitem{832493}
A.~{Feldman} and S.~{Muthukrishnan}.
\newblock Tradeoffs for packet classification.
\newblock In {\em IEEE INFOCOM}, volume~3, pages 1193--1202, 2000.

\bibitem{vpp}
Linux Foundation.
\newblock Vector packet processing (vpp) platform, Oct 2020.

\bibitem{suricata}
Open Information~Security Foundation.
\newblock Suricata - intrusion detection system, nov 2020.

\bibitem{10.1145/1542476.1542528}
Andreas Gal, Brendan Eich, Mike Shaver, David Anderson, David Mandelin,
  Mohammad~R. Haghighat, Blake Kaplan, Graydon Hoare, Boris Zbarsky, Jason
  Orendorff, Jesse Ruderman, Edwin~W. Smith, Rick Reitmaier, Michael Bebenita,
  Mason Chang, and Michael Franz.
\newblock Trace-based {Just-in-Time} type specialization for dynamic languages.
\newblock In {\em Proceedings of the 30th ACM SIGPLAN Conference on Programming
  Language Design and Implementation}, PLDI '09, page 465–478, 2009.

\bibitem{10.1145/1543135.1542528}
Andreas Gal, Brendan Eich, Mike Shaver, David Anderson, David Mandelin,
  Mohammad~R. Haghighat, Blake Kaplan, Graydon Hoare, Boris Zbarsky, Jason
  Orendorff, Jesse Ruderman, Edwin~W. Smith, Rick Reitmaier, Michael Bebenita,
  Mason Chang, and Michael Franz.
\newblock Trace-based just-in-time type specialization for dynamic languages.
\newblock {\em SIGPLAN Not.}, 44(6):465–478, June 2009.

\bibitem{10.5555/2772722.2772728}
Jong~Hun Han, Prashanth Mundkur, Charalampos Rotsos, Gianni Antichi, Nirav~H.
  Dave, Andrew~William Moore, and Peter~G. Neumann.
\newblock {Blueswitch}: Enabling provably consistent configuration of network
  switches.
\newblock In {\em Proceedings of the Eleventh ACM/IEEE Symposium on
  Architectures for Networking and Communications Systems}, ANCS '15, page
  17–27, USA, 2015. IEEE Computer Society.

\bibitem{han2015softnic}
Sangjin Han, Keon Jang, Aurojit Panda, Shoumik Palkar, Dongsu Han, and Sylvia
  Ratnasamy.
\newblock Softnic: A software nic to augment hardware.
\newblock 2015.

\bibitem{hoiland2018express}
Toke H{\o}iland-J{\o}rgensen, Jesper~Dangaard Brouer, Daniel Borkmann, John
  Fastabend, Tom Herbert, David Ahern, and David Miller.
\newblock {The eXpress Data Path: Fast Programmable Packet Processing in the
  Operating System Kernel}.
\newblock In {\em {Proceedings of the 14th International Conference on Emerging
  Networking EXperiments and Technologies}}, CoNEXT '18, pages 54--66, New
  York, NY, USA, 2018. ACM.

\bibitem{10.1145/178243.178478}
Urs H\"{o}lzle and David Ungar.
\newblock Optimizing dynamically-dispatched calls with run-time type feedback.
\newblock In {\em Proceedings of the ACM SIGPLAN 1994 Conference on Programming
  Language Design and Implementation}, PLDI '94, page 326–336, 1994.

\bibitem{hopps:katran}
Christian Hopps.
\newblock Katran: A high performance layer 4 load balancer.
\newblock September 2019.
\newblock \url{https://github.com/facebookincubator/katran}.

\bibitem{google:k8s}
Google Inc.
\newblock {Kubernetes: Production-Grade Container Orchestration}, July 2019.

\bibitem{propellerllvm}
Google Inc.
\newblock Propeller: Profile guided optimizing large scale llvm-based relinker,
  Oct 2019.

\bibitem{iyer2019performance}
Rishabh Iyer, Luis Pedrosa, Arseniy Zaostrovnykh, Solal Pirelli, Katerina
  Argyraki, and George Candea.
\newblock Performance contracts for software network functions.
\newblock In {\em 16th $\{$USENIX$\}$ Symposium on Networked Systems Design and
  Implementation ($\{$NSDI$\}$ 19)}, pages 517--530, 2019.

\bibitem{kubernetes-dualstack}
Joab Jackson.
\newblock Kubernetes long road to dual {IPv4/IPv6} support.
\newblock The New Stack, 2019.

\bibitem{joshi2002denali}
Rajeev Joshi, Greg Nelson, and Keith Randall.
\newblock Denali: A goal-directed superoptimizer.
\newblock {\em ACM SIGPLAN Notices}, 37(5):304--314, 2002.

\bibitem{metron}
Georgios~P. Katsikas, Tom Barbette, Dejan Kosti{\'c}, Rebecca Steinert, and
  Gerald Q.~Maguire Jr.
\newblock Metron: {NFV} service chains at the true speed of the underlying
  hardware.
\newblock In {\em 15th {USENIX} Symposium on Networked Systems Design and
  Implementation ({NSDI} 18)}, pages 171--186, Renton, WA, April 2018. {USENIX}
  Association.

\bibitem{katsikas2016snf}
Georgios~P Katsikas, Marcel Enguehard, Maciej Ku{\'z}niar, Gerald~Q Maguire~Jr,
  and Dejan Kosti{\'c}.
\newblock Snf: Synthesizing high performance nfv service chains.
\newblock {\em PeerJ Computer Science}, 2:e98, 2016.

\bibitem{kazemian2012header}
Peyman Kazemian, George Varghese, and Nick McKeown.
\newblock Header space analysis: Static checking for networks.
\newblock In {\em Presented as part of the 9th $\{$USENIX$\}$ Symposium on
  Networked Systems Design and Implementation ($\{$NSDI$\}$ 12)}, pages
  113--126, 2012.

\bibitem{kazemian2012stanford}
Peyman Kazemian, George Varghese, and Nick McKeown.
\newblock Header space analysis: Static checking for networks.
\newblock In {\em 9th {USENIX} Symposium on Networked Systems Design and
  Implementation ({NSDI} 12)}, pages 113--126, San Jose, CA, April 2012.
  {USENIX} Association.

\bibitem{6379165}
J.~{Kempf}, B.~{Johansson}, S.~{Pettersson}, H.~{Lüning}, and T.~{Nilsson}.
\newblock Moving the mobile {Evolved Packet Core} to the cloud.
\newblock In {\em IEEE International Conference on Wireless and Mobile
  Computing, Networking and Communications (WiMob)}, pages 784--791, 2012.

\bibitem{kohn2018adaptive}
Andr{\'e} Kohn, Viktor Leis, and Thomas Neumann.
\newblock Adaptive execution of compiled queries.
\newblock In {\em 2018 IEEE 34th International Conference on Data Engineering
  (ICDE)}, pages 197--208. IEEE, 2018.

\bibitem{I-D.ietf-quic-manageability}
Mirja Kuehlewind and Brian Trammell.
\newblock Manageability of the {QUIC} transport protocol.
\newblock Internet-Draft draft-ietf-quic-manageability-09, January 2021.

\bibitem{lattner2004llvm}
Chris Lattner and Vikram Adve.
\newblock Llvm: A compilation framework for lifelong program analysis \&
  transformation.
\newblock In {\em International Symposium on Code Generation and Optimization,
  2004. CGO 2004.}, pages 75--86. IEEE, 2004.

\bibitem{procieee_2019}
L.~{Linguaglossa}, S.~{Lange}, S.~{Pontarelli}, G.~{R{\'e}tv{\'a}ri},
  D.~{Rossi}, T.~{Zinner}, R.~{Bifulco}, M.~{Jarschel}, and G.~{Bianchi}.
\newblock Survey of performance acceleration techniques for {Network Function
  Virtualization}.
\newblock {\em Proceedings of the IEEE}, 107(4):746--764, 2019.

\bibitem{liu2018microboxes}
Guyue Liu, Yuxin Ren, Mykola Yurchenko, K.~K. Ramakrishnan, and Timothy Wood.
\newblock Microboxes: High performance nfv with customizable, asynchronous tcp
  stacks and dynamic subscriptions.
\newblock SIGCOMM '18, page 504–517, New York, NY, USA, 2018. Association for
  Computing Machinery.

\bibitem{manousis2020contention}
Antonis Manousis, Rahul~Anand Sharma, Vyas Sekar, and Justine Sherry.
\newblock Contention-aware performance prediction for virtualized network
  functions.
\newblock In {\em Proceedings of the Annual Conference of the ACM Special
  Interest Group on Data Communication on the Applications, Technologies,
  Architectures, and Protocols for Computer Communication}, SIGCOMM '20, page
  270–282, New York, NY, USA, 2020. Association for Computing Machinery.

\bibitem{martins2014clickos}
Joao Martins, Mohamed Ahmed, Costin Raiciu, Vladimir Olteanu, Michio Honda,
  Roberto Bifulco, and Felipe Huici.
\newblock Clickos and the art of network function virtualization.
\newblock In {\em Proceedings of the 11th USENIX Conference on Networked
  Systems Design and Implementation}, NSDI’14, page 459–473, USA, 2014.
  USENIX Association.

\bibitem{miano2019service}
S.~{Miano}, M.~{Bertrone}, F.~{Risso}, M.~V. {Bernal}, Y.~{Lu}, J.~{Pi}, and
  A.~{Shaikh}.
\newblock A service-agnostic software framework for fast and efficient
  in-kernel network services.
\newblock In {\em 2019 ACM/IEEE Symposium on Architectures for Networking and
  Communications Systems (ANCS)}, pages 1--9, 2019.

\bibitem{polycube-tnsm}
S.~{Miano}, F.~{Risso}, M.~V. {Bernal}, M.~{Bertrone}, and Y.~{Lu}.
\newblock A framework for ebpf-based network functions in an era of
  microservices.
\newblock {\em IEEE Transactions on Network and Service Management}, pages
  1--1, 2021.

\bibitem{miano2019securing}
Sebastiano Miano, Matteo Bertrone, Fulvio Risso, Mauricio~V\'{a}squez Bernal,
  Yunsong Lu, and Jianwen Pi.
\newblock Securing linux with a faster and scalable iptables.
\newblock {\em SIGCOMM Comput. Commun. Rev.}, 49(3):2–17, November 2019.

\bibitem{miano2018creating}
Sebastiano Miano, Matteo Bertrone, Fulvio Risso, Massimo Tumolo, and
  Mauricio~V{\'a}squez Bernal.
\newblock Creating complex network services with ebpf: Experience and lessons
  learned.
\newblock In {\em 2018 IEEE 19th International Conference on High Performance
  Switching and Routing (HPSR)}, pages 1--8. IEEE, 2018.

\bibitem{molnar2016dataplane}
L\'{a}szl\'{o} Moln\'{a}r, Gergely Pongr\'{a}cz, G\'{a}bor Enyedi,
  Zolt\'{a}n~Lajos Kis, Levente Csikor, Ferenc Juh\'{a}sz, Attila
  K\H{o}r\"{o}si, and G\'{a}bor R\'{e}tv\'{a}ri.
\newblock Dataplane specialization for high-performance openflow software
  switching.
\newblock In {\em Proceedings of the 2016 ACM SIGCOMM Conference}, SIGCOMM '16,
  page 539–552, New York, NY, USA, 2016. Association for Computing Machinery.

\bibitem{10.1145/3428245}
Manasij Mukherjee, Pranav Kant, Zhengyang Liu, and John Regehr.
\newblock Dataflow-based pruning for speeding up superoptimization.
\newblock {\em Proc. ACM Program. Lang.}, 4(OOPSLA), November 2020.

\bibitem{olteanu2018beamer}
Vladimir Olteanu, Alexandru Agache, Andrei Voinescu, and Costin Raiciu.
\newblock Stateless datacenter load-balancing with beamer.
\newblock In {\em 15th {USENIX} Symposium on Networked Systems Design and
  Implementation ({NSDI} 18)}, pages 125--139, Renton, WA, April 2018. {USENIX}
  Association.

\bibitem{java-tweak}
Oracle.
\newblock {Java HotSpot VM Options}, 2021.
\newblock
  \url{https://www.oracle.com/java/technologies/javase/vmoptions-jsp.html}.

\bibitem{ovn-geneve}
The {Open Virtual Network} architecture: {Tunnel} encapsulations.
\newblock Open vSwitch Manual, 2018.

\bibitem{panchenko2019bolt}
Maksim Panchenko, Rafael Auler, Bill Nell, and Guilherme Ottoni.
\newblock Bolt: a practical binary optimizer for data centers and beyond.
\newblock In {\em Proceedings of the 2019 IEEE/ACM International Symposium on
  Code Generation and Optimization}, pages 2--14. IEEE Press, 2019.

\bibitem{panda2016netbricks}
Aurojit Panda, Sangjin Han, Keon Jang, Melvin Walls, Sylvia Ratnasamy, and
  Scott Shenker.
\newblock Netbricks: Taking the v out of nfv.
\newblock In {\em Proceedings of the 12th USENIX Conference on Operating
  Systems Design and Implementation}, OSDI’16, page 203–216, USA, 2016.
  USENIX Association.

\bibitem{deng2018redundant}
Bangwen Pedrosa and Wenfei Wu.
\newblock Redundant logic elimination in network functions.
\newblock In {\em Proceedings of the ACM SIGCOMM 2018 Conference on Posters and
  Demos}, pages 78--80. ACM, 2018.

\bibitem{pedrosa2018automated}
Luis Pedrosa, Rishabh Iyer, Arseniy Zaostrovnykh, Jonas Fietz, and Katerina
  Argyraki.
\newblock Automated synthesis of adversarial workloads for network functions.
\newblock In {\em Proceedings of the 2018 Conference of the ACM Special
  Interest Group on Data Communication}, SIGCOMM ’18, page 372–385, New
  York, NY, USA, 2018. Association for Computing Machinery.

\bibitem{openvswitch2015}
Ben Pfaff, Justin Pettit, Teemu Koponen, Ethan~J. Jackson, Andy Zhou, Jarno
  Rajahalme, Jesse Gross, Alex Wang, Jonathan Stringer, Pravin Shelar, Keith
  Amidon, and Mart\'{\i}n Casado.
\newblock {The Design and Implementation of Open vSwitch}.
\newblock In {\em {Proceedings of the 12th USENIX Conference on Networked
  Systems Design and Implementation}}, NSDI'15, pages 117--130. USENIX
  Association, 2015.

\bibitem{10.1145/2872362.2872387}
Phitchaya~Mangpo Phothilimthana, Aditya Thakur, Rastislav Bodik, and Dinakar
  Dhurjati.
\newblock Scaling up superoptimization.
\newblock In {\em Proceedings of the Twenty-First International Conference on
  Architectural Support for Programming Languages and Operating Systems},
  ASPLOS '16, page 297–310, New York, NY, USA, 2016. Association for
  Computing Machinery.

\bibitem{gcc}
GNU Project.
\newblock Gnu compiler collection.

\bibitem{rath2017symperf}
Felix Rath, Johannes Krude, Jan R{\"u}th, Daniel Schemmel, Oliver Hohlfeld,
  J{\'o}~{\'A} Bitsch, and Klaus Wehrle.
\newblock Symperf: Predicting network function performance.
\newblock In {\em Proceedings of the SIGCOMM Posters and Demos}, pages 34--36.
  ACM, 2017.

\bibitem{retvari2017dynamic}
G{\'a}bor R{\'e}tv{\'a}ri, L{\'a}szl{\'o} Moln{\'a}r, G{\'a}bor Enyedi, and
  Gergely Pongr{\'a}cz.
\newblock Dynamic compilation and optimization of packet processing programs.
\newblock {\em ACM SIGCOMM NetPL}, 2017.

\bibitem{rizzo12}
Luigi Rizzo.
\newblock {netmap: A Novel Framework for Fast Packet I/O}.
\newblock In {\em Annual Technical Conference (ATC)}. USENIX Association, 2012.

\bibitem{souper}
Raimondas Sasnauskas, Yang Chen, Peter Collingbourne, Jeroen Ketema, Jubi
  Taneja, and John Regehr.
\newblock Souper: A synthesizing superoptimizer.
\newblock 2017.

\bibitem{sekar2012design}
Vyas Sekar, Norbert Egi, Sylvia Ratnasamy, Michael~K. Reiter, and Guangyu Shi.
\newblock Design and implementation of a consolidated middlebox architecture.
\newblock In {\em Presented as part of the 9th {USENIX} Symposium on Networked
  Systems Design and Implementation ({NSDI} 12)}, pages 323--336, San Jose, CA,
  2012. {USENIX}.

\bibitem{180672}
Vyas Sekar, Norbert Egi, Sylvia Ratnasamy, Michael~K. Reiter, and Guangyu Shi.
\newblock Design and implementation of a consolidated middlebox architecture.
\newblock In {\em 9th {USENIX} Symposium on Networked Systems Design and
  Implementation ({NSDI} 12)}, pages 323--336, San Jose, CA, April 2012.
  {USENIX} Association.

\bibitem{10.1145/2774993.2775000}
Muhammad Shahbaz and Nick Feamster.
\newblock The case for an intermediate representation for programmable data
  planes.
\newblock In {\em Proceedings of the 1st ACM SIGCOMM Symposium on Software
  Defined Networking Research}, SOSR '15, 2015.

\bibitem{bpf_map_batch}
Yonghong Song.
\newblock bpf: adding map batch processing support, Aug 2019.

\bibitem{snort}
Sourcefire.
\newblock {Snort - Network Intrusion Detection \& Prevention System}, nov 2020.

\bibitem{stack-java}
StackOverflow.
\newblock What can cause my code to run slower when the server {JIT} is
  activated?, 2011.
\newblock
  \url{https://stackoverflow.com/questions/2923989/what-can-cause-my-code-to-run-slower-when-the-server-jit-is-activated}.

\bibitem{10.1145/3098822.3098826}
Chen Sun, Jun Bi, Zhilong Zheng, Heng Yu, and Hongxin Hu.
\newblock Nfp: Enabling network function parallelism in nfv.
\newblock In {\em Proceedings of the Conference of the ACM Special Interest
  Group on Data Communication}, SIGCOMM '17, page 43–56, New York, NY, USA,
  2017. Association for Computing Machinery.

\bibitem{taylor2007classbench}
David~E Taylor and Jonathan~S Turner.
\newblock Classbench: A packet classification benchmark.
\newblock {\em IEEE/ACM transactions on networking}, 15(3):499--511, 2007.

\bibitem{wintermeyer20}
Patrick Wintermeyer, Maria Apostolaki, Alexander Dietm\"{u}ller, and Laurent
  Vanbever.
\newblock {P2GO: P4 Profile-Guided Optimizations}.
\newblock In {\em Hot Topics in Networks (HotNets)}. ACM, 2020.

\bibitem{guard-elision}
Jonathan Worthington.
\newblock Eliminating unrequired guards.
\newblock 6guts, 2018.

\bibitem{cloudflare:unimog}
David Wragg.
\newblock {Unimog - Cloudflare’s edge load balancer}.
\newblock sep 2020.

\bibitem{xhonneux2018leveraging}
Mathieu Xhonneux, Fabien Duchene, and Olivier Bonaventure.
\newblock Leveraging ebpf for programmable network functions with ipv6 segment
  routing.
\newblock In {\em Proceedings of the 14th International Conference on Emerging
  Networking EXperiments and Technologies}, CoNEXT ’18, page 67–72, New
  York, NY, USA, 2018. Association for Computing Machinery.

\bibitem{zhang2012micro}
Rui Zhang, Saumya Debray, and Richard~T Snodgrass.
\newblock Micro-specialization: dynamic code specialization of database
  management systems.
\newblock In {\em Proceedings of the Tenth International Symposium on Code
  Generation and Optimization}, pages 63--73, 2012.

\end{thebibliography}

\end{document}